\documentclass[a4paper]{article}

\usepackage{amsmath,amsfonts,proof}
\usepackage{hyperref,url,natbib}

\newcommand{\mytitle}{The pitfalls of verifying floating-point computations}

\title{\mytitle}

\author{David Monniaux\\CNRS / \'Ecole normale sup\'erieure%
\footnote{The author is now at CNRS / VERIMAG, Grenoble.}}

\renewcommand{\cite}[2][]{\citep[#1]{#2}}

\newcommand{\titlea}{\maketitle}
\newcommand{\titleb}{}

\begin{document}
\newcommand{\elast}{\varepsilon_{\text{last}}}
\newcommand{\eabs}{\varepsilon_{\text{abs}}}
\newcommand{\erel}{\varepsilon_{\text{rel}}}
\newcommand{\emin}{\varepsilon_{\min}}
\newcommand{\eden}{e_{\text{den}}}
\newcommand{\nmant}{p-1}
\newcommand{\bbR}{\mathbb{R}}
\newcommand{\bbQ}{\mathbb{Q}}

\newcommand{\band}[2]{#1 \wedge #2}
\newcommand{\bor}[2]{#1 \vee #2}
\newcommand{\bneg}[1]{\neg #1}
\newcommand{\btrue}{\textsf{true}}
\newcommand{\bfalse}{\textsf{false}}
\newcommand{\subst}[3]{#1 [ #2 \mapsto #3 ]}
\newcommand{\instrText}[1]{\textsf{#1}}
\newcommand{\nop}{\instrText{skip}}
\newcommand{\assign}[2]{#1 := #2}
\newcommand{\seq}[2]{#1; #2}
\newcommand{\cifthenelse}[3]{\instrText{if~} #1 \instrText{~then~} #2 \instrText{~else~} #3}
\newcommand{\HoareTriple}[3]{\{#1\} #2 \{#3\}}
\newcommand{\ndt}{\textsf{ndt}}

\newcommand{\astree}{\textsc{Astr\'ee}}
\newcommand{\TM}{$^{\text{TM}}$}

\newenvironment{norm}
               {\list{}{\listparindent 1.5em%
                        \itemindent    1.5em%
                        \leftmargin 1.5em%
                        \rightmargin   1.5em}%
                \item\relax\em}
               {\endlist}

\titlea

\begin{abstract}
Current critical systems often use a lot of floating-point
computations, and thus the testing or static analysis of programs
containing floating-point operators has become a priority. However,
correctly defining the semantics of common implementations of
floating-point is tricky, because semantics may change according to many
factors beyond source-code level, such as choices made by
compilers. We here give concrete examples of problems that can appear
and solutions for implementing in analysis software.
\end{abstract}

\titleb


\sloppy
\section{Introduction}
In the past, critical applications often used fixed-point
computations. However, with the wide availability of processors with
hardware floating-point units, many current critical applications
(say, for controlling automotive or aerospace systems) use
floating-point operations. Such applications have to undergo stringent
testing or validation. In this paper, we show how the particularities
of floating-point implementations can hinder testing methodologies
and have to be cared for in techniques based on program semantics, be
them assisted proofs, or automatic analysers.

It has been known for a long time that it was erroneous to compute
with floating-point numbers and operations as though they were on the
real field. There exist entire treatises discussing the topic of
stability in numerical algorithms from the point of view of the
applied mathematician: whether or not some algorithm, when implemented
with floating-point, will give ``good'' approximations of the real
result; we will however not discuss such issues in this paper.
The purpose of this paper is to show the kind of
difficulties that floating-point computations pose for static analysis
and program testing methods: both for
defining the semantics of the programs to be analysed,
and for defining and implementing the analysis techniques.
We shall not discuss ``bugs'' in floating-point implementations in
microprocessors,%
\footnote{William Kahan has interesting cases, see his short paper
\emph{Beastly Numbers}.} but, rather, how misunderstandings and
non-intuitive behaviours of correct hardware implementations
affect the safety of programs and program
analysis techniques.

Many of the issues that we discuss here are known to floating-point
arithmetic experts. However, they have often been ignored or
misunderstood by designers of programming languages, compilers,
verification techniques and program analysers. Some of them were taken
care of, at least partially, in the definition of Java and
the latest standard for~C,
as well as modern hardware implementations. Our primary
objective, however, is to educate the designers of verification
systems, who in general do not have the luxury to change the hardware
implementation, the programming language, the compilers used or the
algorithms of the systems that they have to verify. Some of the issues
we discuss pertain to the Intel{\TM} 386, i86, Pentium{\TM} lines (IA32{\TM}
architecture), which will most probably continue to be used for
embedded work for a number of years. Because of this focus on program
analysis for critical systems, we shall  be particularly interested in
programs written in the C programming language, because this language
is often favoured for embedded systems. We shall in particular discuss
some implications of the most recent standard of that language, ``C99''
\cite{C99}. In order to emphasise that the issues we discuss are real,
we shall give specific program fragments as well as versions of
compilers, libraries and operating systems with which ``surprising''
behaviours were witnessed.

All current major microprocessor architectures (IA32, x86\_64,
PowerPC) support IEEE-754 floating-point arithmetic \cite{IEEE-754},
now also an international standard~\cite{IEC-60559}; microcontrollers
based on these architectures also do so. Other microcontrollers
and microprocessors often implement some variant of it with some
features omitted. The specification for the default floating-point
arithmetic in the Java programming language is also based on
IEEE-754. For these reasons, we shall focus on this standard and its
main implementations, though some of our remarks apply to other kinds
of systems. For the sake of a better understanding,
in Section~\ref{part:IEEE-754}, we recall the basics of IEEE-754
arithmetic.

Despite, or perhaps because of, the prevalence of ``IEEE-compliant''
systems, there exist a number of myths of what IEEE-compliance really
entails from the point of view of program semantics. We shall discuss
the following myths, among others:
\begin{itemize}
\item ``Since C's \texttt{float} (resp. \texttt{double})
  type is mapped to IEEE single (resp. double) precision arithmetic,
  arithmetic operations have a uniquely defined meaning across
  platforms.''
\item ``Arithmetic operations are deterministic; that is, if I do
  \texttt{z=x+y} in two places in the same program and my program
  never touches \texttt{x} and \texttt{y} in the meantime,
  then the results should be the same.''
\item A variant:
  ``If \texttt{x < 1} tests true at one point, then \texttt{x < 1}
  stays true later if I never modify \texttt{x}.''
\item ``The same program, strictly compliant with the C standard with no
  ``undefined behaviours'', should yield identical results if compiled
  on the same IEEE-compliant platform by different compliant compilers.''
\end{itemize}

A well-known cause for such unintuitive discrepancies is the 80-bit
internal floating-point registers on the Intel
platform.~\cite[Appendix~D]{Sun-numeric} In Section~\ref{part:IA32},
we shall expand on such issues
and show, for instance, how low-level issues such as register
allocation \cite[chapter~11]{Appel_modern_compiler_C_97}
and the insertion of logging instructions with no
``apparent'' computational effects can change the final results of
computations. In Section~\ref{part:PowerPC} we shall discuss issues
pertaining to the PowerPC architecture.

An important factor throughout the discussion is that it is not the
hardware platform that matters in itself, but its combination
with the software context, including the compiler, libraries,
and possible run-time environment.
Compatibility has to be
appreciated at the level of the application writer --- how code
written using types mapped to IEEE normalised formats will effectively
behave when compiled and run.
Indeed, the standard recalls \cite[\S 1.1]{IEC-60559,IEEE-754}:
\begin{norm}
It is the environment the
programmer or user of the system sees that conforms or fails to conform to
this standard. Hardware components that require software support to
conform shall not be said to conform apart from such software.
\end{norm}

IEEE-754 standardises a few basic operations; however, many programs use
functions such as sine, cosine, \dots, which are not specified by this
standard and are generally not strictly specified in the system
documentation.  In Section~\ref{part:library}, we shall explore
some difficulties with respect to mathematical libraries.
In addition to issues related to certain floating-point
implementations, or certain mathematical libraries, there are issues
more particularly related to the C programming language, its compilers
and libraries. Section~\ref{part:compiler} explores such
system dependencies. Section~\ref{part:input-output} explores issues
with input and output of floating-point values.

A commonly held opinion is that whatever the discrepancies, they will be
negligible enough and should not have noticeable consequences.
In Section~\ref{part:example}, we give a complete example of some
seemingly innocuous floating-point code fragment based on real-life
industrial code.
We illustrate how the floating-point ``oddities''
that we explained in the preceding sections can lead to rare and
extremely hard to diagnose run-time errors.

While the focus of the paper is on the C programming language, in
Section~\ref{part:Java} we shall discuss a few aspects of the
compilation of Java, a language reputedly more ``predictable'', which
many advocate for use in embedded systems.

While in most of the paper we dismiss some incorrectly optimistic
beliefs about how floating-point computations behave and what one can
safely suppose about them for program analysis, an opposite
misconception exists: that floating-point is inherently so complex and
tricky that it is impossible to do any kind of sound analysis, or do
any kind of sound reasoning, on programs using floating-point, except
perhaps in very simple cases. By sound analysis
(Sec.~\ref{part:semantic-bases}), we mean that when one
analyses the program in order to prove properties (say, ``variable $x$
does not exceed $42$), using some automated, semi-automated or manual
proof techniques, then the results that are obtained truly hold of the
concrete system (e.g. one does not prove the above statement when in
reality $x$ can reach $42.000000001$).
The {\astree} static analyser%
\footnote{{\astree} is a static analyser specialised on a subset of C
  suitable for many critical applications. It features specific, sound
  and precise analyses for floating-point digital filters. It was
  successfully applied to several classes of critical code, especially
  fly-by-wire software. See \url{http://www.astree.ens.fr} as well as
  \cite{BlanchetCousotEtAl02-NJ,BlanchetCousotEtAl_PLDI03,ASTREE_ESOP05}.
  Simply put, {\astree} takes as input the source code of a program,
  a specification of bounds on program inputs, and computes,
  symbolically, a super-set of reachable program states and possible
  program executions, from which it extracts properties interesting to
  the user, and proves that certain undesirable behaviours (overflow,
  array access out of bounds, bad pointer dereference, violation of
  assertion) are impossible.}
implements mathematically sound analyses by
taking into account some ``error bounds'' derived from the
specification of the concrete floating-point semantics. The existence
of {\astree} and its use in an industrial context demonstrate that
it is possible to obtain sound analyses with a reasonable cost, at
least for some classes of applications. We shall therefore refer to this
tool as a practical example further on.

In Section~\ref{part:implications} we analyse the consequences of
the issues discussed in previous sections
on abstract in\-ter\-pre\-ta\-tion-based
static analysis and other validation techniques, and show how to
obtain sound results.

\section{IEEE-754: a reminder}\label{part:IEEE-754}
All current general-purpose microprocessors, and many
microcontrollers, implement hardware floating-point as a variant of
standard ANSI/IEEE-754 \cite{IEEE-754}, later adopted as international standard
IEC-60559 \cite{IEC-60559}. We thus begin by an overview of this
standard. We shall not, however, describe how algorithms can make
clever use of the standard for numerical computation purposes, and
refer the reader to papers and presentations
from e.g. William Kahan%
\footnote{\url{http://www.cs.berkeley.edu/~wkahan/}}
on such issues.

\subsection{Numbers}
\label{part:hexadecimal}
IEEE floating point numbers are of the following kinds:
\begin{description}
\item[Zeroes] There exist both a $+0$ and a $-0$. Most operations
  behave identically regardless of the sign of the zero, however
  one gets different results if one
  extracts the sign bit from the number, or if one divides
  a nonzero number by a zero (the sign of the zero and that of the
  nonzero other operand determine whether $+\infty$ or $-\infty$ is returned).

  Many programs do not depend on this feature of IEEE arithmetic and
  would behave identically if only a single zero was provided. This is
  the case of most programs implementing real (as opposed to complex)
  computations. However, discriminating between $+0$ and $-0$ allows
  for some simpler implementations of certain classes of algorithms.

  An important example of the use for separate $\pm 0$ is
  complex arithmetic~\cite{Kahan_branch_cuts}: if branch
  cuts are located along the real or complex axes, then distinguishing
  $+0$ and $-0$ allow making functions continuous from both sides of the
  slit, while having a single zero value introduces a discontinuity.
  As an example, consider $\Im \log(z)$ with $z=x+iy$, $x < 0$:
  $\lim_{y \rightarrow 0^+} \Im \log(x+iy) = +\pi$ and
  $\lim_{y \rightarrow 0^-} \Im \log(x+iy) = -\pi$, and thus it makes
  sense to define by continuity from both sides:
  $\log(x+0i)=+\pi$ and $\log(x-0i)=-\pi$. This behaviour of complex
  arithmetic along branch cuts is mandated
  by the C99 standard~\cite[\S 7.3.3]{C99}.

\item[Infinities] Infinities are generated by divisions by zero or by
  \emph{overflow} (computations of numbers of such a large magnitude
  that they cannot be represented).
\item[NaNs] The special values \emph{Not a Number} (NaN) represent the
  result of operations that cannot have a meaningful result in terms
  of a finite number or infinity. Such is for instance the case of
  $(+\infty) - (+\infty)$, $0/0$ or $\sqrt{-1}$.
\item[Normal numbers] (Also known as \emph{normalised numbers}.)
  These are the most common kind of nonzero
  representable reals.
\item[Subnormal numbers] (Also known as \emph{denormalised numbers}
  or \emph{denormals}.)
  These represent some values very close to
  zero. They pose special issues regarding rounding errors.
\end{description}

IEEE specifies 5 possible kinds of exceptions. These exceptions can,
at the request of the programmer, be substituted for ``silent''
default responses:
\begin{description}
\item[Invalid operation] This is the exception corresponding to the
  conditions producing a NaN as silent response.
\item[Overflow] This exception is raised when the result of an
  operation is a number too large in magnitude to be represented in
  the data type. The silent response is to generate
  $\pm\infty$.
\item[Division by zero]  Raised by $x/\pm 0$ where $x \neq 0$. The
  default response is to produce $\pm\infty$, depending on the sign of
  $x$ and of $\pm 0$.
\item[Underflow] This exception is raised when a result is too small
  in magnitude to be computed accurately. This is generally harmless
  in programs; however, care must be taken when computing error bounds
  on floating-point computations, because error bounds on underflow
  differ from those on normal computations (see below).
\item[Inexact] The ``real'' result of a computation cannot be exactly
  represented by a floating-point number. The silent response is to
  round the number, which is a behaviour that the vast majority of
  programs using floating-point numbers rely upon. However, rounding
  has to be correctly taking into account for sound analysis.
\end{description}

In typical critical embedded critical programs, invalid operations,
divisions by zero, and overflows are undesirable conditions. In such
systems, inputs vary within controlled ranges, many variables model
physical quantities (which also should lie within some reasonable
range), and thus NaNs and $\pm\infty$ should not appear. This is
especially true since infinities may generate NaNs down the line
(e.g. $(+\infty)-(+\infty)=NaN$), NaNs propagate throughout the
algorithms, and converting a NaN to an output quantity (conversion to
integer after scaling, for instance) yields an undefined result.
For this reason, in most critical embedded systems, the generation of
a NaN or infinity is an undesirable condition, which the designers of
the system will want to rule out, including through formal methods
(see Sec.~\ref{part:implications}).
The invalid operation, overflow
and division by zero exceptions may be activated, so that they are
raised as soon as an undesirable result is computed. They may trigger
the termination of the program, or the running of a ``degraded''
version of it with limited functionality or complexity. This is why
one of the features of analysers such as {\astree} is to detect where
in programs such exceptions may be raised. In order to achieve this
goal, the analyser will have to bound variables and results of
floating-point operations, which requires some sound analysis techniques and
motivates this paper.

Floating point numbers are represented as follows:
$x = \pm s.m$ where $1 \leq m < 2$ is the \emph{mantissa} or
\emph{significand}, which
has a fixed number $p$ of bits, and $s =
2^e$ the \emph{scaling factor} ($E_{\min} \leq e \leq E_{\max}$
is the \emph{exponent}).
The difference introduced by changing the last binary digit of the
mantissa is $\pm s.\elast$ where $\elast = 2^{-(p-1)}$:
the \emph{unit in the last place} or \emph{ulp}.
Such a decomposition is unique for a given number
if we impose that the leftmost digit of the
mantissa is $1$ --- this is called a \emph{normalised representation}.
Except in the case of numbers of very small magnitude, IEEE-754 always
works with normalised representations.

The IEEE-754 \emph{single precision} type, generally associated with C's
\texttt{float} type \cite[\S F.2]{C99}, has $p=24$, $E_{\min}=-126$,
$E_{\max}=+127$. The IEEE-754 \emph{single precision} type, associated to C's
\texttt{double}, has $p=53$, $E_{\min}=-1022$,
$E_{\max}=+1023$.

We thus obtain normalised floating-point representations of the form:
\begin{equation}
x = \pm {\left[1.m_1 \ldots m_{\nmant}\right]}_2 . 2^e,
\end{equation}
noting $[vvv]_2$ the representation of a number in terms of binary
digits $vvv$.

Conversion to and from decimal representation is delicate; special
care must be taken in order not to introduce inaccuracies
or discrepancies.~\cite{SteeleWhite90,Clinger90}. Because of this,
C99 introduces hexadecimal floating-point literals in source
  code.~\cite[\S 6.4.4.2]{C99} Their syntax is as follows:
  \texttt{0x\textit{mmmmmm.mmmm}p$\pm$\textit{ee}} where
  \texttt{\textit{mmmmmm.mmmm}} is a mantissa in hexadecimal, possibly
  containing a point, and \texttt{\textit{ee}} is an exponent, written
  in decimal, possibly preceded by a sign. They are interpreted as
  $[\texttt{\textit{mmmmmm.mmmm}}]_{16} \times
  2^{\texttt{\textit{ee}}}$.
Hexadecimal floating-point representations are especially important
when values must be represented exactly, for reproducible results ---
for instance, for testing ``borderline cases'' in algorithms. For this
reason, we shall use them in this paper wherever it is important to
specify exact values.
See also Section~\ref{part:input-output} for more information on
inputting and outputting floating-point values.

\subsection{Rounding}\label{part:rounding}
Each real number $x$ is mapped to a
floating-point value $r(x)$ by a uniquely defined rounding function;
the choice of this function is determined by the \emph{rounding mode}.

Generally, floating-point units and libraries allow the program to
change the current rounding mode; C99 mandates \texttt{fegetround()} and
\texttt{fesetround()} to respectively get and set the current rounding
mode on IEEE-compliant platforms. Other platforms (BSD etc.) provide
\texttt{fpgetround()} and \texttt{fpsetround()}. Finally,
on some other platforms, it may be necessary to use assembly language
instructions that directly change the mode bits inside the floating-point
unit.%
\footnote{The {\astree} static analyser, which uses directed rounding
  internally as described in Sec.~\ref{part:analysers}, contains a
  module that strives to provide a way of changing rounding modes on
  most current, common platforms. We regret the uneven support of the
  standardised functions.}
IEEE-754 mandates four standard rounding modes:
\begin{itemize}
\item Round-to-$+\infty$ (directed rounding to $+\infty$):
  $r(x)$ is the least floating
  point value greater than or equal to $x$.
\item Round-to-$-\infty$ (directed rounding to $-\infty$):
  $r(x)$ is the greatest floating
  point value smaller than or equal to $x$.
\item Round-to-$0$: $r(x)$ is the floating-point value of the same
  sign as $x$ such that $|r(x)|$ is the greatest floating
  point value smaller than or equal to $|x|$.
\item Round-to-nearest: $r(x)$ is the floating-point value closest to
  $x$ with the usual distance; if two floating-point value are equally
  close to~$x$, then $r(x)$ is the one whose least significant bit is
  equal to zero. Rounding to nearest with different precisions twice
  in a row (say, first to double precision, then to single precision)
  may yield different results than rounding directly to the final
  type; this is known as \emph{double rounding}.
  Some implications of double-rounding are investigated in
  Sec.~\ref{part:double-rounding}.
\end{itemize}

The default mode is round-to-nearest, and this is the only one used in the
vast majority of programs.

The discussion on floating-point errors frequently refers to the
notion of ``unit in the last place'' or ``ulp''; but there are
different definitions of this notion, with subtle
differences~\cite{MullerULP}.
If the range of exponents were unbounded, there would exist a positive real
$\erel$ such that, for all~$x$, $|x - r(x)| \leq \erel.|x|$.
This \emph{relative error} characterisation is used in some papers
analysing the impact of roundoff errors. It is, however, incorrect in
the context of IEEE-754, where the range of exponents is bounded:
not only there is the possibility of overflow, but
there exists a least positive representable number, which introduces
an \emph{absolute error} for values very close to
zero~\cite[\S 7.4.3]{Mine_PhD}\cite{Mine_ESOP04}.
As a result, for any floating-point type with a bounded number of bits,
there exist two positive reals $\erel$ and $\eabs$ such that
\begin{equation}
|x - r(x)| \leq \max(\erel.|x|, \eabs)\label{eqn:rounding1}
\end{equation}

The following, coarser, property may be easier to use in some contexts:
\begin{equation}
|x - r(x)| \leq \erel.|x| +\eabs\label{eqn:rounding2}
\end{equation}

\subsection{Operations}\label{part:operations}
IEEE-754 standardises 5 operations: addition (which we shall note $\oplus$
in order to distinguish it from the operation over the reals), subtraction
($\ominus$), multiplication ($\otimes$), division ($\oslash$), and
also square root.

IEEE-754 specifies \emph{exact rounding} \cite[\S 1.5]{Goldberg91}: the
result of a floating-point operation is the same as if the operation
were performed on the real numbers with the given inputs, then rounded
according to the rules in the preceding section. Thus,
$x \oplus y$ is defined as $r(x + y)$, with $x$ and $y$ taken as
elements of $\bbR \cup \{ -\infty, +\infty \}$; the same applies for
the other operators.

One difficulty, when reasoning about floating-point computations,
both with human and automated reasoning, is that floating-point
operations ``almost'' behave like their name-sakes on the reals, but
not quite exactly.
For instance, it is well-known that floating-point operations are \emph{not
  associative} (e.g
$(10^{20} \oplus 1) \allowbreak \ominus \allowbreak 10^{20}
\allowbreak = \allowbreak 0 \allowbreak \neq \allowbreak 1 =
(10^{20} \ominus 10^{20}) \allowbreak \oplus \allowbreak 1$ using
IEEE-754 double precision operations). Many
symbolic computation techniques, used in compiler optimisers, program
analysers or proof assistants,
assume some good algebraic properties of the
arithmetic in order to be sound; thus, if applied directly and
straightforwardly on floating-point operation, they are unsound.
Yet, some compilers rely on such properties to perform optimisations
(see~\ref{part:optimisations}).
In Section~\ref{part:analysers}, we shall explain how it is possible to
make such analysis methods sound with respect to floating-point.

\section{Architecture-dependent issues}
Some commonplace architectures, or, more appropriately, some
commonplace ways of compiling programs and using floating-point
programs on certain commonplace architectures, can introduce subtle
inconsistencies between program executions.

\subsection{IA32, x86\_64 architectures}\label{part:IA32}
The IA32 architecture, originating from Intel, encompasses processors
such as the i386, i486 and the various Pentium variants. It was until
recently the most common architecture for personal computers, but has
since been superseded for that usage by the x86\_64 architecture,
a 64-bit extension of IA32.%
\footnote{%
  AMD made a 64-bit architecture called AMD64{\TM}.
  Intel then produced a mostly compatible architecture, calling it
  Intel{\textregistered}~64 (not to be confused with IA64{\TM},
  another, incompatible, 64-bit architecture from Intel, found
  in the Itanium{\TM} processor),
  after briefly calling it EM64T{\TM}. Microsoft Windows documentation
  calls these architectures x64. We chose a middle ground and
  followed the terminology of GNU/Linux distributions (x86\_64).}
IA32, however, is a very commonplace architecture for embedded systems,
including with embedded variants of the i386 and i486 processors.
IA32 offers almost complete upward compatibility from the 8086 processor,
first released in 1978; it features a floating-point unit, often nicknamed
x87, mostly upwardly compatible with the 8087 coprocessor, first released in
1980.

Later, another floating-point unit, known as SSE,
was added to the architecture, with full support for IEEE-754 starting
from the Pentium~4 processor; it is now the preferred unit to use.
The use of the x87 unit is deprecated on x86\_64 ---
for instance, the popular \texttt{gcc} compiler
does not use it by default on this platform, and the documentation for
Microsoft Visual Studio C++ on x86\_64 calls this unit
``legacy floating-point''.
However, microcontrollers and embedded microprocessors are
likely not to include this SSE unit in the years to come, even if they
include x87 hardware floating-point.

\subsubsection{x87 floating-point unit}\label{part:x87}
Processors of the IA32 architecture (Intel 386, 486,
Pentium etc. and compatibles) feature a floating-point unit often
known as ``x87''~\cite[chapter 8]{IA32-1}.
\begin{norm}
It supports the floating-point, integer, and packed BCD integer data types and the floating-point processing algorithms and exception handling architecture defined in the IEEE Standard 754 for Binary Floating-Point Arithmetic.
\end{norm}
This unit has 80-bit registers in  ``double extended''
format (64-bit mantissa and 15-bit exponent), often associated to the
\texttt{long double} C type; IEEE-754 specifies the
possibility of such a format. The unit can
read and write data to memory in this 80-bit format or in standard
IEEE-754 single and double precision.

By default, all
operations performed on CPU registers are done with 64-bit precision,
but it is possible to reduce precision to 24-bit (same as IEEE single
precision) and 53-bit (same as IEEE double precision) mantissas by
setting some bits in the unit's control
register.\cite[\S 8.1.5.2]{IA32-1}
These precision
settings, however, do not affect the range of exponents available, and only
affect a limited subset of operations (containing all operations
specified in IEEE-754). As a result, changing these precisions
settings will not result in floating-point operations being performed
in strict IEEE-754 single or double precision.%
\footnote{%
By ``strict IEEE-754 behaviour'', ``strict IEEE-754 single precision'' or
``strict IEEE-754 double precision'', we mean that each individual
basic arithmetic operation is performed as if the computation were
done over the real numbers, then the result rounded to single or
double precision.}

The most usual way of generating code for the IA32 is to hold
temporaries --- and, in optimised code, program variables --- in the x87
registers. Doing so yields more compact and efficient code than always
storing register values into memory and reloading them. However, it is
not always possible to do everything inside registers, and compilers
then generally spill extra temporary values to main memory,~
\cite[chapter~11]{Appel_modern_compiler_C_97}
using the format associated to the type associated to the value by the typing
rules of the language. For instance, a \texttt{double} temporary will
be spilt to a 64-bit double precision memory cell. This means
that \emph{the final result of the computations depend on how the compiler
allocates registers}, since temporaries (and possibly variables)
will incur or not incur rounding whether or not they are spilt to main
memory.

As an example, the following program compiled
with \texttt{gcc} 4.0.1 \cite{gcc} under Linux will print $10^{308}$
(\texttt{1E308}):
\begin{verbatim}
  double v = 1E308;
  double x = (v * v) / v;
  printf("%g %d\n", x, x==v);
\end{verbatim}
How is that possible? \texttt{v * v} done in double precision will
overflow, and thus yield $+\infty$, and the final result should be
$+\infty$. However, since all computations are performed in extended
precision, with a larger exponent range, the computations do not
overflow. If we use the
\texttt{-ffloat-store} option, which forces \texttt{gcc} to store
floating-point variables in memory, we obtain $+\infty$.

The result of computations can actually depend on compilation options
or compiler versions, or anything that affects propagation. With the
same compiler and system, the following program prints $10^{308}$
(when compiled in optimised mode (\texttt{-O}), while it prints
$+\infty$ when compiled in default mode.
\begin{verbatim}
double foo(double v) {
        double y = v * v;
        return (y / v);
}

main() { printf("%g\n", foo(1E308));}
\end{verbatim}
Examination of the assembly
code shows that when optimising, the compiler reuses the value of
\texttt{y} stored in a register, while it saves and reloads \texttt{y}
to and from main memory in non-optimised mode.

A common optimisation is \emph{inlining} --- that is, replacing a call
to a function by the expansion of the code of the function at the
point of call. For simple functions (such as small arithmetic
operations, e.g. $x \mapsto x^2$), this can increase performance
significantly, since function calls induce costs (saving registers,
passing parameters, performing the call, handling return values). C99
\cite[\S 6.7.4]{C99} and C++ have an \texttt{inline} keyword in order
to pinpoint functions that should be inlined (however, compilers are
free to inline or not to inline such functions; they may also inline
other functions when it is safe to do so). However, on x87, whether or
not inlining is performed may change the semantics of the code!

Consider what \texttt{gcc} 4.0.1 on IA32 does with the following program,
depending on whether the optimisation switch \texttt{-O} is passed:
\begin{verbatim}
static inline double f(double x) {
  return x/1E308;
}

double square(double x) {
  double y = x*x;
  return y;
}

int main(void) {
  printf("%g\n", f(square(1E308)));
}
\end{verbatim}
\texttt{gcc} does not inline functions when optimisation is turned
off. The \texttt{square} function returns a \texttt{double}, but the
calling convention is to return floating point values into a x87
register --- thus in \texttt{long double} format. Thus, when 
\texttt{square} is called, it returns approximately $10^{716}$, which
fits in  \texttt{long double}  but not \texttt{double} format. But
when \texttt{f} is called, the parameter is passed on the stack
--- thus as a  \texttt{double}, $+\infty$. The program therefore
prints $+\infty$. In comparison, if the program is compiled with
optimisation on, \texttt{f} is inlined; no parameter passing takes
place, thus no conversion to \texttt{double} before division, and thus
the final result printed is $10^{308}$.

It is somewhat common for beginners to add a comparison check to $0$ before
computing a division, in order to avoid possible division-by-zero
exceptions or the generation of infinite results. A first objection to
this practise is that, anyway, computing
$1/x$ for $x$ very close to zero will generate
very large numbers that will most probably result in overflows later. Indeed,
programmers lacking experience with floating-point are advised that
they should hardly ever use strict comparison tests
($=$, $\neq$, $<$ and $>$ as opposed to $\leq$ and $\geq$) with
floating-point operands, as it is somewhat pointless to exclude
some singularity point by excluding one single value,
since it will anyway be surrounded by
mathematical instability or at least very large values which will
break the algorithms.
Another objection, which few programmers know about and that we wish
to draw attention to,
is that it may actually fail to work, depending on what the
compiler does --- that is, the program may actually test that
$\texttt{x} \neq 0$, then, further down, find that $\texttt{x} = 0$
without any apparent change to \texttt{x}. How can this be possible?

Consider the following source code (see Section~\ref{part:hexadecimal} for the meaning of hexadecimal floating-point constants):%
\begin{verbatim}
/* zero_nonzero.c */
void do_nothing(double *x) { }

int main(void) {
  double x = 0x1p-1022, y = 0x1p100, z;
  do_nothing(&y);
  z = x / y;
  if (z != 0) {
    do_nothing(&z);
    assert(z != 0);
  }
}
\end{verbatim}

This program exhibits different behaviours depending on various
factors, even when one uses the same compiler (\texttt{gcc} version
4.0.2 on IA32):
\begin{itemize}
\item If it is compiled without optimisation, \texttt{x / y} is computed
  as a \texttt{long double} then converted into a IEEE-754 double precision
  number ($0$) in order to be saved into memory variable \texttt{z},
  which is then reloaded from memory for the test.
  The \texttt{if} statement is thus not taken.
\item If it is compiled as a single source code with optimisation,
  \texttt{gcc} performs some kind of global analysis which understands
  that \verb+do_nothing+ does nothing. Then, it does constant
  propagation, sees that \texttt{z} is $0$, thus that the \texttt{if}
  statement is not taken, and finally that \texttt{main()} performs no
  side effect. It then effectively compiles \texttt{main()} as a ``no
  operation''.
\item If it is compiled  as two source codes (one for each function)
  with optimisation,
  \texttt{gcc} is not able to use information about what
  \verb+do_nothing()+ does when compiling \verb+main()+. It will thus
  generate two function calls to \verb+do_nothing()+, and will not
  assume that the value of \texttt{y} (respectively, \texttt{z}) is
  conserved across  \verb+do_nothing(&y)+ (respectively,
  \verb+do_nothing(&z)+).
  The \verb+z != 0+ test
  is performed on a nonzero \texttt{long double} quantity and thus the
  test is taken. However, after the \verb+do_nothing(&z)+ function call,
  \texttt{z} is reloaded from main memory as the value $0$ (because
  conversion to double-precision flushed it to~$0$).
  As a consequence, the final assertion fails, somehow contrary to
  what many programmers would expect.
\item With the same compilation setup as the last case, removing
  \verb+do_nothing(&z)+ results in the assertion being true:
  \texttt{z} is then not flushed to memory and thus kept as an
  extended precision nonzero floating-point value.
\end{itemize}

One should therefore be extra careful with strict comparisons, because
the comparison may be performed on extended precision values, and fail
to hold later after the values have been converted to single or double
precision --- which may happen or not depending on a variety of
factors including compiler optimisations and ``no-operation''
statements.

We are surprised by these discrepancies. After all, the C specification says
\cite[5.1.2.3, \emph{program execution}, \S 12, ex.~4]{C99}:
\begin{norm}
Implementations employing wide registers have to take care to honour appropriate
semantics. Values are independent of whether they are represented in a register or in memory. For example, an implicit spilling of a register is not permitted to alter the value. Also, an explicit store and load is required to round to the precision of the storage type.
\end{norm}
However, this paragraph, being an example, is not
normative.~\cite[foreword, \S 6]{C99}. By reading the C specification
more generally, one gets the impression that such hidden side-effects
(``hidden'', that is, not corresponding to program statements) are
prohibited.

The above examples indicate that common debugging practises that
apparently should not change the computational semantics may
actually alter the result of computations. Adding a logging statement
in the middle of a computation may alter the scheduling of registers,
for instance by forcing some value to be spilt into main memory and
thus undergo additional rounding. As an example, simply inserting a
\verb+printf("%g\n", y);+ call after the computation of \texttt{y} in
the above \texttt{square} function forces \texttt{y} to be spilt to memory, and
thus the final result then becomes $+\infty$ regardless of
optimisation. Similarly, our \verb+do_nothing()+ function may be
understood as a place-holder for logging or debugging statements which
are not supposed to affect the state of the variables.

In addition, it is commonplace to disable
optimisation when one intends to use a software debugger, because in
optimised code, the compiled code corresponding to distinct statements
may become fused, variables may not reside in a well-defined location,
etc. However, as we have seen, simply disabling or enabling
optimisation may change computational results.

\subsubsection{Double rounding}\label{part:double-rounding}
In some circumstances, floating-point results are rounded twice in a
row, first to a type $A$ then to a type $B$. Surprisingly, such
\emph{double rounding} can yield different results from direct
rounding to the destination type.%
\footnote{This problem has been known for a
long time.\cite[chapter~6]{FigueroaPhD}\cite[4.2]{Goldberg91}}
Such is the case, for instance, of results computed in the \texttt{long
double} 80-bit type of the x87 floating-point registers, then
rounded to the IEEE double precision type for storage in memory.
In round-to-$0$, round-to-$+\infty$ and round-to-$-\infty$ modes, this is
not a problem provided that the values representable by type $B$ are a
subset of those representable by type $A$. However, in
round-to-nearest mode, there exist some borderline cases where
differences are exhibited.

In order to define the round-to-nearest mode, one has to define
arbitrarily how to round a real exactly in the middle between the
nearest floating-point values. IEEE-754 chooses round-to-even
\cite[\S 4.1]{IEC-60559,IEEE-754}:%
\footnote{\cite[1.5]{Goldberg91} explains a rationale for this.}
\begin{norm}
In this mode, the representable value nearest to the infinitely precise result shall be delivered; if the two nearest representable values are equally near, the one with its least significant bit equal to zero shall be delivered.
\end{norm}

This definition makes it possible for double rounding to yield
different results than single rounding to the destination
type. Consider a floating-point type $B$ where two consecutive values
are $x_0$ and $x_0 + \delta_B$, and another floating-type $A$ containing
all values in $B$ and also $x_0 + \delta_B/2$. There exists
$\delta_A > 0$
such that all reals in the interval
$I = (x_0 + \delta_B/2 - \delta_A/2, x_0 + \delta_B/2)$
get rounded to $x_0 + \delta_B/2$ when mapped to type~$A$.
We shall suppose that the mantissa of $x_0$ finishes by a
$1$.
If $x \in I$, then indirect rounding yields:
$x \rightarrow_A (x_0 + \delta_B/2) \rightarrow_B (x_0 + \delta_B)$ and
direct rounding yields: $x \rightarrow_B x_0$.

A practical example is with $x_0 = 1+2^{-52}$, $\delta=2^{-52}$ and
$r = x_0 + y$ where $y = (\delta/2) (1 - 2^{-11})$. Both $x_0$ and $y$
are exactly representable in IEEE-754 double precision ($B$).
\begin{verbatim}
  double x0 = 0x1.0000000000001p0;
  double y = 0x1p-53 * (1. - 0x1p-11);
  double z1 = x0 + y;
  double z2 = (long double) x0 + (long double) y;
  printf("%a %a\n", z1, z2);
\end{verbatim}

In order to get strict IEEE-754 double
precision computations for the \texttt{double} type,
we execute double-precision
computations on the SSE unit (see Section~\ref{part:SSE}) of an x86\_64 or
Pentium~4 processor.
We then obtain that $z_1 = x_0$ and that $z_2 = x_0 + 2^{-52}$: both
$z_1$ and $z_2$ are \texttt{double} values obtained by apparently
identical computations (the sum of $x_0$ and $y$), but the value that
$z_2$ holds will have
been doubly rounded (first to extended precision, then to double
precision for storing to $z_2$) while $z_1$ holds a value directly
rounded to double precision.

Note that, on IA32, depending on compilation modes, the above
discrepancy may disappear, with both values undergoing double
rounding: on IA32 Linux / \texttt{gcc}-4.0.1 with default options,
the computations on \texttt{double} will
be actually performed in the \texttt{long double} type inside
the x87 unit, then converted to IEEE double precision. There is thus
no difference between the formulae computing $z_1$ and $z_2$.

Another example was reported as a bug by a user who noticed
inconsistencies between 387 and SSE but did not identify the source of
the problem:
\begin{verbatim}
  double b = 0x1.fffffffffffffp-1;
  double x = 1 / b;
\end{verbatim}
Let $\varepsilon=2^{-53}$, then $\texttt{b}=1-\varepsilon$.
$1/b = 1+\varepsilon+\varepsilon^2+\dots$; the nearest double precision
numbers are $1$ and $1+2\varepsilon$ and thus direct rounding gives
$\texttt{x}=1+2\varepsilon$. However, rounding to extended precision will
give $1+\varepsilon$, which is rounded to $1$ when converting to double
precision.

A similar problem occurs with rounding behaviour near infinities: see
the definition of round-to-nearest for large values \cite[\S 4.1]{IEEE-754}:
\begin{norm}
However, an infinitely precise result with magnitude at least
$2^{E_{\text{max}}} (2 - 2^{-p})$ shall round to $\infty$ with no change in sign.
\end{norm}
For IEEE double-precision, $E_{\text{max}}=1023$ and $p=53$; let us take $x_0$ to
be the greatest representable real,
$M_{\text{double}}=2^{E_{\text{max}}}(2-2^{-(p-1)})$
and $y = 2^{970} (1 - 2^{-11})$. With a similar program as above,
$r=x_0+y$ gets rounded to $z_1=x_0$ in IEEE double precision, but gets
rounded to $2^{E_{\text{max}}}(2-2^{-p})$ in extended precision. As a
result, the subsequent conversion into IEEE double precision will
yield $+\infty$.

Double rounding can also cause some subtle differences for very small
numbers that are rounded into subnormal double-precision values if
computed in IEEE-754 double precision: if one uses the
``double-precision'' mode of the x87 FPU, these numbers will be
rounded into normalised values inside the FPU register, because of a
wider range of negative exponents; then they will be rounded again
into double-precision subnormals when written to memory. This is known
as \emph{double-rounding on underflow}. Working around double-rounding
on underflow is somewhat tedious
(however, the phenomenon is exhibited by $\times$ and $/$,
not by $+$ and $-$).~\cite{JavaGrandeSC98}
A concrete example~: taking
\begin{eqnarray*}
\texttt{x} = \texttt{0x1.8000000000001p-1018}\approx 5.34018\times10^{-307}\\
\texttt{y} = \texttt{0x1.0000000000001p+56}\approx7.20576\times10^{16},
\end{eqnarray*}
then $\texttt{x} \oslash \texttt{y} = \texttt{0x0.0000000000001p-1022}$
in strict IEEE-754 double precision and
$\texttt{x} \oslash \texttt{y} = \texttt{0x0.0000000000002p-1022}$
with the x87 in ``double precision mode''.

\subsubsection{SSE floating-point unit}\label{part:SSE}
Intel introduced the SSE floating-point
unit \cite[chapter 10]{IA32-1} in the Pentium~III processor,
then the SSE2 extension in the Pentium~4
\cite[chapter~11]{IA32-1}. These extensions to the x86 instruction set
contain, respectively, IEEE-compatible single-precision and
double-precision instructions.%
\footnote{In addition to scalar instructions, SSE and SSE2 introduce
  various vector instructions, that is, instructions operating over
  several operands in parallel. We shall see in
  \S\ref{part:optimisations} that compilers may rearrange expressions
  incorrectly in order to take advantage of these vector
  instructions. For now, we only discuss the scalar instructions.}
One can make \texttt{gcc} generate code for the SSE subsystem with the
\texttt{-mfpmath=sse} option; since SSE is only available for certain
processors, it is also necessary to specify, for instance,
\texttt{-march=pentium4}. On x86\_64, \texttt{-mfpmath=sse} is the
default, but \texttt{-mfpmath=387} forces the use of the x87 unit.

Note the implication: the same program may give different results when
compiled on 32-bit and 64-bit ``PCs'' (or even the same machine,
depending on whether one compiles in 32-bit or 64-bit mode) because of
the difference in the default floating-point subsystem used.

The differences may seem academic, but the following incident proves
they are not. The Objective Caml system \cite{OCaml} includes two
compilers: one compiles Caml code into a portable bytecode, executed
by a virtual machine (this virtual machine is written in~C);
the other one compiles directly to native assembly code. One user of
Objective Caml on the recent Intel-based Apple Macintosh computers
complained of a mysterious ``bug'' to the Caml maintainers: the same
program gave slightly different results across the bytecode and
native code versions. The problem could not be reproduced on other
platforms, such as Linux on the same kind of processors. It turned out
that, as all Intel-based Macintosh machines have a SSE unit, Apple
configured \texttt{gcc} to use the SSE unit by default. As a
consequence, on this platform, by default,
the Objective Caml virtual machine uses the SSE unit when
executing bytecode performing floating-point computations: for
instance, an isolated floating-point addition will be performed as a
sequence of two loads from double-precision operands, addition with
the result in a double-precision register, and then save to a
double-precision variable. The native
code compiler, however, uses the x87 unit:
the same floating-point addition is thus performed as a
sequence of two loads from double-precision operands, addition with
the result in an extended-precision register, and then save to a
double-precision variable. As we pointed out in the preceding section,
these two sequences of operations are not equivalent in the default
round-to-nearest mode, due to double rounding. It turned out that the
user had stumbled upon a value resulting in double rounding. The
widespread lack of awareness of floating-point issues resulted in the
user blaming the discrepancy on a bug in Objective~Caml!

In addition, the SSE unit offers some non-IEEE-754 compliant modes for
better efficiency: with the \emph{flush-to-zero} flag
\cite[\S 10.2.3.3]{IA32-1} on, subnormals are not generated and are
replaced by zeroes; this is more efficient. As we noted in
Section~\ref{part:rounding}, this does not hamper obtaining good bounds
on the errors introduced by floating-point computations; also, we can
assume the worst-case situation and suppose that this flag is on when
we derive error bounds.

The flush-to-zero flag, however, has another notable consequence:
$x \ominus y = 0$ is no longer equivalent to $x = y$. As an example,
if $x = 2^{-1022}$ and $y = 1.5 \times 2^{-1022}$, then
$y \ominus x = 2^{-1023}$ in normal mode, and $y \ominus x = 0$ in
flush-to-zero mode. Analysers should therefore be careful when
replacing comparisons by ``equivalent'' comparisons.

In addition, there exists a \emph{denormals-are-zero} flag \cite[\S 10.2.3.4]{IA32-1}: if it is on, all subnormal operands are considered to be zero, which improves performance. It is still possible to obtain bounds on the errors of
floating point computations by assuming that operands are offset by an
amount of at most $\pm 2^{e_{\min}-(\nmant)}$ before being computed
upon. However, techniques based on exact replays of instruction
sequences will have to replay the sequence with the same value of the flag.

\subsubsection{Problems and solutions}
We have shown that computations on the \texttt{float}
(respectively, \texttt{double}) types are not performed on the x87 unit
exactly as if each atomic arithmetic operation were performed
with IEEE-754 single (respectively, double precision) operands and
result, and that what is actually performed may depend on seemingly
irrelevant factors such as calls to tracing or printing functions.
This goes against the widespread myth that the result of the
evaluation of a floating-point expression should be the same across
all platforms compatible with IEEE-754.

This discrepancy has long been
known to some people in the programming language community, and some
``fixes'' have been proposed.
For instance, \texttt{gcc} has a
\verb+-ffloat-store+ option, flushing floating-point variables to
memory.~\cite{gcc}
Indeed, the \texttt{gcc} manual
\cite{gcc} says:
\begin{norm}
     On 68000 and x86 systems, for instance, you can get paradoxical
     results if you test the precise values of floating point numbers.
     For example, you can find that a floating point value which is not
     a NaN is not equal to itself.  This results from the fact that the
     floating point registers hold a few more bits of precision than
     fit in a \texttt{double} in memory.  Compiled code moves values between
     memory and floating point registers at its convenience, and moving
     them into memory truncates them. 
     You can partially avoid this problem by using the \texttt{-ffloat-store}
     option.
\end{norm}
The manual refers to the following option:
\begin{norm}
\texttt{-ffloat-store}
     Do not store floating point variables in registers, and inhibit
     other options that might change whether a floating point value is
     taken from a register or memory.

     This option prevents undesirable excess precision on machines
     [\dots] where the floating registers [\dots] keep more
     precision than a `double' is supposed to have.  Similarly for the
     x86 architecture.  For most programs, the excess precision does
     only good, but a few programs rely on the precise definition of
     IEEE floating point.[sic]  Use `\texttt{-ffloat-store}' for such
     programs, after
     modifying them to store all pertinent intermediate computations
     into variables.
\end{norm}
Note that this option does not force unnamed temporaries to be flushed
to memory, as shown by experiments. To our knowledge, no compiler
offers the choice to always spill temporaries to memory, or to flush
temporaries to \texttt{long double} memory, which would at least
remove the worst problem, which is the non-reproducibility of results
depending on factors independent of the computation code
(register allocation differences caused by compiler options or
debugging code, etc.). We suggest that compilers should include such
options.

Unfortunately, anyway, systematically flushing values to single- or
double-precision memory cells
do not reconstitute strict IEEE-754 single- or double- precision
rounding behaviour in round-to-nearest mode,
because of the double rounding problem (\ref{part:double-rounding}).
In addition, the \texttt{-ffloat-store} option is difficult to use,
because it only affects
program variables and not temporaries: to approximate strict IEEE-754
behaviour, the programmer would have to rewrite all program formulae
to store temporaries in variables. This does not seem to be reasonable
for human-written code, but may be possible with automatically
generated code --- it is frequent that control/command applications
are implemented in a high-level language such as Simulink%
\footnote{Simulink\TM is a tool for modelling dynamic systems and control
applications, using e.g. networks of numeric filters. The control part
may then be compiled to hardware or software.\\
\url{http://www.mathworks.com/products/simulink/}}
Lustre \cite{LUSTRE} or Scade\TM,%
\footnote{Scade is a commercial product based on LUSTRE.\\
\url{http://www.esterel-technologies.com/products/scade-suite/}}
then compiled into~C~\cite{C99}.

Another possibility is to force the floating-point unit to limit
the width of the mantissa to that of IEEE-754 basic formats (single
or double precision).%
\footnote{This is the default setting on FreeBSD~4, presumably in
  order to achieve closer compatibility with strict IEEE-754 single
  or double precision computations.}
This \emph{mostly} solves the double-rounding problem.
However, there is no way to
constrain the range of the exponents, and thus these modes do not
allow exact simulation of strict computations on IEEE single and double
precision formats, when overflows and underflows are possible.
For instance, the \texttt{square} program of Sec.~\ref{part:x87},
which results in an overflow to $+\infty$
if computations are done on IEEE double precision numbers,
does not result in overflows if run with the x87 in double
precision mode. Let us note, however, that if a computation never
results in overflows or underflows when done with IEEE-754 double-precision
(resp. single-) arithmetic, it can be
exactly simulated with the x87 in double-precision (resp. single) mode.%
\footnote{Let us consider the round-to-nearest case.
If $|r| \leq M_{\text{double}}$ (where $M_{\text{double}}$ is the
greatest representable double precision number), then the x87 in
  double-precision mode rounds exactly like IEEE-754 double-precision
  arithmetic.
If $M_{\text{double}} < |r| < 2^{E_{\text{max}}} (2 - 2^{-p})$, then,
according to the round-to-nearest rules (including ``round-to-even''
for $r = 2^{E_{\text{max}}} (2 - 2^{-p})$), $r$ is rounded to $\pm
M_{\text{double}}$ on the x87, which is correct with respect to
IEEE-754. If $|r| \geq 2^{E_{\text{max}}} (2 - 2^{-p})$, then rounding
$r$ results in an overflow. The cases for the other rounding modes are
simpler.}

If one wants semantics almost exactly
faithful to strict IEEE-754 single or double precision computations in
round-to-nearest mode,
including with respect to overflow and underflow conditions,
one can use, at the same
time, limitation of precision and options and programming style
that force operands to be systematically written to memory between
floating-point operations. This incurs some performance
loss; furthermore, there will still be slight
discrepancy due to double rounding on underflow.
A simpler solution for current personal computers is simply to force
the compiler to use the SSE unit for computations on IEEE-754 types;
however, most embedded systems using IA32 microprocessors or
microcontrollers do not use processors equipped with this unit.

Yet another solution is to do all computations in \texttt{long double} format. This solves all inconsistency issues. However, \texttt{long double} formats are not the same between processors and operating systems, thus this workaround leads to portability problems.

\subsection{PowerPC architecture}\label{part:PowerPC}
The floating point operations implemented in the PowerPC architecture
are compatible with IEEE-754 \cite[\S1.2.2.3, \S 3.2]{PowerPC-32}. However,
\cite[\S 4.2.2]{PowerPC-32} also points out that:
\begin{norm}
The architecture supports the IEEE-754 floating-point standard, but requires software support to conform with that standard.
\end{norm}

The PowerPC architecture features floating-point
multiply-add instructions \cite[\S 4.2.2.2]{PowerPC-32}. These perform
$(a,b,c) \mapsto \pm a.b \pm c$ computations in one instruction ---
with obvious benefits for computations such as matrix computations
\cite[\S 26.1]{Cormen},
dot products, or Horner's rule for evaluating polynomials \cite[\S
32.1]{Cormen}.
Note, however, that they are not semantically
equivalent to performing separate addition, multiplication and
optional negate IEEE-compatible instructions; in fact, intermediate
results are computed with extra precision \cite[D.2]{PowerPC-32}.
Whether these instructions are used or not depends on the compiler,
optimisation options, and also how the compiler subdivides
instructions. For instance, \texttt{gcc} 3.3 compiles the following
code using the multiply-add instruction if optimisation (\texttt{-O})
is turned on, but without it if optimisation is off, yielding
different semantics:%
\footnote{\texttt{gcc} has an option \texttt{-mno-fused-madd} to turn
  off the use of this instruction.}
\begin{verbatim}
double dotProduct(double a1, double b1,
                  double a2, double b2) {
        return a1*b1 + a2*b2;
}
\end{verbatim} 

In addition, the \texttt{fpscr} control register has a \texttt{NI}
bit, which, if on, possibly enables implementation-dependent semantics
different from IEEE-754 semantics.~\cite[\S 2.1.4]{PowerPC-32}.
This alternate semantics may include behaviours disallowed by IEEE-754
regardless of which data formats and precisions are used.
For instance, on the MPC750 family, such non-compliant behaviour
encompasses flushing subnormal results to zero, rounding subnormal
operands to zero, and treating NaNs differently
\cite[\S 2.2.4]{MPC750}. Similar caveats apply as in Section~\ref{part:SSE}.

\section{Mathematical functions}\label{part:library}
Many operations related to floating-point are not implemented in
hardware; most programs using floating-point will thus rely on
suitable support libraries. Our purpose, in this section, is not to
comprehensively list bugs in current floating-point libraries; it is
to illustrate, using examples from common operating systems and
runtime environments, the kind of problems that may happen.

\subsection{Transcendental functions}
\label{part:transcendental}
A first remark is that, though IEEE-754
specifies the behaviour of elementary
operations $+$, $-$, $\times$, $/$ and $\sqrt{}$, it does not specify
the behaviour of other functions, including the popular trigonometric
functions. These are generally supplied by a system library,
occasionally by the hardware.

As an example, consider the sine function. On the x87, it is
implemented in hardware; on Linux IA32, the GNU libc function
\texttt{sin()} is just a wrapper around the hardware call, or,
depending on compilation options, can be replaced by inline assembly.%
\footnote{The latter behaviour is triggered by option
  \texttt{-ffast-math}. The documentation for this function says
     \emph{it
     can result in incorrect output for programs which depend on an
     exact implementation of IEEE or ISO rules/specifications for math
     functions}.}
Intel \cite[\S 8.3.10]{IA32-1} and AMD \cite[\S 6.5.5]{AMD64-1} claim that
their transcendental instructions (on recent processors)
commit errors less than 1~ulp in
round-to-nearest mode; however it is to be understood that this is
after the operands of the trigonometric functions are reduced modulo
$2\pi$, which is done currently using a 66-bit approximation
for~$\pi$.~\cite[\S 8.3.8]{IA32-1}
However, the AMD-K5 used up to 256 bits for approximating $\pi/2$.~\cite{AMD_K5_transcendental}

One obtains different results for
\verb@sin(0x1969869861.p+0)@ on PCs running Linux. The Intel Pentium~4,
AMD Athlon64 and AMD Opteron processors, and GNU libc
running in 32-bit mode on IA32 or x86\_64
all yield \texttt{-0x1.95b011554d4b5p-1},
However, Mathematica, Sun Sparc
under Solaris and Linux, and  GNU libc on x86\_64 (in 64-bit mode)
yield \texttt{-0x1.95b0115490ca6p-1}.
 
A more striking example of discrepancies is
$\sin(p)$ where $p = 14885392687$. This value was chosen so that
$\sin(p)$ is close to $0$, in order to demonstrate the impact of
imprecise reduction modulo $2\pi$.%
\footnote{
Consider a rational approximation of $\pi$, i.e. integers
$p$ and $q$ such that $p/q \approx \pi$ (such an approximation can be
obtained by a continued fraction development of $\pi$
\cite{ContinuedFraction}). $\sin(p) \approx \sin(q\pi)=0$.
If $p'$, the result of reduction modulo $2\pi$ of $p$,
is imprecise by a margin of $\varepsilon$ ($\exists k~p'-p = \varepsilon +
2k\pi$), then $\sin(p')-\sin(p) \approx \varepsilon$ ($\sin(x) \sim x$
close to $0$). Such inputs are thus good candidates to illustrate possible lack of precision in the algorithm for reduction modulo  $2\pi$.}
Both the Pentium~4 x87 and Mathematica yield a result
$\sin(p) \approx 1.671 \times 10^{-10}$. However, GNU libc on x86\_64
yields $\sin(p) \approx 1.4798 \times 10^{-10}$, about 11.5\% off!




Note, also, that different processors within the same architecture
can implement the same transcendental functions with different
accuracies. We already noted the difference between the AMD-K5 and the
K6 and following processors with respect to angular reduction.
Intel also notes that the algorithms' precision was improved between
the 80387 / i486DX processors and the Pentium
processors.~\cite[\S 7.5.10]{Pentium-1}
\begin{norm}
With the Intel486 processor and Intel 387 math coprocessor, the
worst-case, transcendental function error is typically 3 or 3.5~ulps,
but is sometimes as large as 4.5~ulps.
\end{norm}
There thus may be floating-point discrepancies between the current
Intel embedded processors (based on i386 / i387) and the current
Pentium workstation processors.

To summarise, one should not expect consistent behaviour of
transcendental functions across libraries, processor manufacturers or
models, although recent developments such as the MPFR library%
\footnote{MPFR, available from \url{http://www.mpfr.org}, is a library built on top of the popular GNU~MP multiprecision library. MPFR provides arbitrary precision floating-point arithmetic with the same four rounding modes as IEEE-754. In addition to the basic arithmetic operations specified by IEEE-754, MPFR also provides various arbitrary precision transcendental functions with guaranteed rounding.}
provide
exactly rounded transcendental functions. In any case, static analysis
tools should never assume that the libraries on the analysis host
platform behave identically to those on the target platform, nor,
unless it is specified in the documentation, that they fulfil
properties such as monotonicity (on intervals where the corresponding
mathematical function is monotonic).

\subsection{Non-default rounding modes}\label{part:buggy_rounding_modes}
We also have found that floating-point libraries
are often poorly tested in ``uncommon'' usage conditions, such as
rounding modes different from ``round-to-nearest''; or, perhaps, that they
are not supposed to work in such modes, but that this fact is not reflected
adequately in documentation.
This is especially of interest for implementers of static analysers
(Section~\ref{part:analysers}), since some form of interval arithmetic
will almost certainly be used in such software.

FreeBSD 4.4 provides the \texttt{fpsetround()} function to set the
rounding mode of the processor. This function is the BSD counterpart of
the C99 \texttt{fesetround()} function. However,
the \texttt{printf()} standard printing function of the 
C library does not work properly if the processor is set in
round-to-$+\infty$ mode: when one attempts to print very large values
(such as $10^{308}$), one can get garbage output (such as a colon
in a location where a digit should be, e.g. \texttt{:e+307}). 

On GNU libc 2.2.93 on IA32 processors, the \texttt{fesetround()}
function only changed the rounding mode of the x87 FPU, while the
\texttt{gcc} compiler also offered the possibility of compiling for
SSE.

On GNU libc 2.3.3 on x86\_64, computing $x^y$ using the \texttt{pow()} function
in round-to-$+\infty$ mode can result in a segmentation violation for
certain values of $x$ and $y$, e.g.
$x = \texttt{0x1.3d027p+6}$ and $y=\texttt{0x1.555p-2}$. As for the
\texttt{exp()} exponential function, it gives a result close to
$2^{502}$ on input $1$, and a negative result on input
\texttt{0x1.75p+0}. The problems were corrected in version 2.3.5.

\subsection{Compiler issues}
\label{part:compiler}
The C standard is fairly restrictive with respect to what compilers
are allowed or not allowed to do with floating-point
operations. Unfortunately, some compilers may fail to comply with the
standard. This may create discrepancies between what really happens on
the one hand, and what users and static
analysis tools expect on the other hand.

\subsubsection{Standard pragmas}
The C standard, by default, allows the compiler some substantial
leeway in the way that floating-point expressions may be
evaluated. While outright simplifications based on operator
associativity are not permitted, since these can be very unsound on
floating-point types \cite[\S 5.1.2.3 \#13]{C99}, the compiler is for
instance permitted to replace a complex expression by a simpler one,
for instance using compound operators (e.g. the fused multiply-and-add
in Section~\ref{part:PowerPC}) \cite[6.5]{C99}:
\begin{norm}
A floating expression may be contracted, that is, evaluated as though
it were an atomic operation, thereby omitting rounding errors implied
by the source code and the expression evaluation method. [\dots]
This license is specifically intended to allow implementations to
exploit fast machine instructions that combine multiple C
operators. As contractions potentially undermine predictability, and
can even decrease accuracy for containing expressions, their use needs
to be well-defined and clearly documented.
\end{norm}
While some desirable properties of contracted expressions \cite[\S
F.6]{C99} are requested, no precise behaviour is made compulsory.

Because of the inconveniences that discrepancies can create,
the standard also mandates a
special directive, \verb+#PRAGMA+ \verb+STDC+ \verb+FP_CONTRACT+
\cite[\S 7.12.2]{C99}, for controlling whether or not such
contractions can be performed. Unfortunately, while many compilers will
contract expressions if they can, few compilers implement this
pragma. As of 2007, \texttt{gcc} (v4.1.1)
ignores the pragma with a warning,
and Microsoft's Visual C++ handles it as a recent addition.

We have explained how, on certain processors such as the x87
(Section~\ref{part:x87}), it was possible to change the precision of
results by setting special flags --- while no access to such flags is
mandated by the C norm, the possibility of various precision modes is
acknowledged by the norm \cite[F.7.2]{C99}.
Furthermore, IEEE-754 mandates the
availability of various rounding modes (Section~\ref{part:rounding}); in
addition, some processors offer further flags that change the
behaviour of floating-point computations.

All changes of modes are done through library functions (or
inline assembly) executed at runtime; at the same time, the C compiler may do
some computations at compile time, without regard to how these modes are set.
\begin{norm}
During translation the IEC~60559 default modes are
in effect: The rounding direction mode is rounding to nearest.
The rounding precision mode (if supported) is set so that results are not
shortened.   Trapping or stopping (if supported) is disabled on all
floating-point exceptions. [\dots] The implementation
should produce a diagnostic message for each translation-time
floating-point exception, other than  inexact; the
implementation should then proceed with the translation of the
program.
\end{norm}

In addition, programs to be compiled
may test or change the floating-point status or
operating modes using library functions, or even inline assembly. If
the compiler performs code reorganisations, then some results may end
up being computed before the applicable rounding modes are set.
For this reason, the C norm introduces \verb+#pragma STDC FENV_ACCESS ON/OFF+
\cite[\S 7.6.1]{C99}:
\begin{norm}
The \texttt{FENV\_ACCESS} pragma provides a means to inform the
implementation when a program might access the floating-point
environment to test floating-point status flags or run under
non-default floating-point control modes. [\dots]  If part of a program
tests floating-point status flags, sets floating-point control modes,
or runs under non-default mode settings, but was translated with the
state for the \texttt{FENV\_ACCESS} pragma off, the behaviour is
undefined. The default state ( on or off) for the pragma is
implementation-defined. [\dots] The purpose of the \texttt{FENV\_ACCESS}
pragma is to allow certain optimisations that could subvert flag tests and mode
changes (e.g., global common subexpression elimination, code motion,
and constant folding). In general, if the state of
\texttt{FENV\_ACCESS} is off,
the translator can assume that default modes are in effect and the
flags are not tested.
\end{norm}

Another effect of this pragma is to change how much the compiler can
evaluate at compile time regarding constant
initialisations.~\cite[F.7.4, F.7.5]{C99}. If it is set to \texttt{OFF},
the compiler can evaluate floating-point constants at compile time,
whereas if they had been evaluated at runtime, they would have
resulted in different values (because of different rounding modes) or
floating-point exception. If it is set to \texttt{ON}, the compiler
may do so only for static constants --- which are generally all
evaluated at compile time and stored as a block of constants in the
binary code of the program.

Unfortunately, as per the preceding pragma, most compilers do not
recognise this pragma. There may, though, be some compilation options
that have some of the same effect. Again, the user should carefully
read the compiler documentation.

\subsubsection{Optimisations and associativity}
\label{part:optimisations}

Some optimising compilers will apply rules such as
associativity, which may significantly alter the outcome of an
algorithm, and thus are \emph{not} allowed to apply according to
the language standard. \cite[\S 5.1.2.3]{C99}

A particularly interesting application of such permissiveness is 
\emph{vectorisation}; that is,
using the features of certain processors (such as IA32, x86\_64 or EM64T
processors, with a SSE unit) that enable doing the same mathematical
operation on several operands in a single instruction. Take for
instance the following program, which sums an array:
\begin{verbatim}
double s = 0.0;
for(int i=0; i<n; i++) {
  s = s + t[i];
}
\end{verbatim}

This code is not immediately vectorisable. However,
assuming that $n$ is a multiple of two \emph{and that addition is
associative}, one may rewrite this code as follows:
\begin{verbatim}
double sa[2], s; sa[0]=sa[1]= 0.0;
for(int i=0; i<n/2; i++) {
  sa[0] = sa[0] + t[i*2+0];
  sa[1] = sa[1] + t[i*2+1];
}
s = sa[0] + sa[1];
\end{verbatim}
That is, we sum separately the elements of the array with even or odd
indexes. Depending on other conditions such as memory alignment, this code may
be immediately vectorised on processors that can do two simultaneous
double precision operations with one single vector instruction.

If we compile the first code fragment above using Intel's \texttt{icc} compiler with
options \verb@-xW -O2@ (optimise for Pentium~4 and compatible
processors), we see that the loop has been automatically vectorised
into something resembling the second fragment. This is because, by
default, \texttt{icc} uses a ``relaxed'' conformance mode with respect
to floating-point; if one specifies \verb@-fp-model precise@, then the
application of associativity is prohibited and the compiler says
``loop was not vectorised: modifying order of reduction not allowed
under given switches''. Beta versions of \texttt{gcc}~4.2 do not
vectorise this loop when using \verb@-O3 -ftree-vectorize@,
but they will do so on appropriate
platforms with option \verb@-ffast-math@ or the aptly named
\verb@-funsafe-math-optimisations@.

Another class of optimisations involves the assumption, by the
compiler, that certain values ($\pm\infty$, NaN) do not occur in
regular computations, or that the distinction between $\pm 0$ does not
matter. This reflects the needs of most computations, but may be
inappropriate in some contexts; for instance, some computations on
complex numbers use the fact that $\pm 0$ are
distinct~\cite{Kahan_branch_cuts}. It is thus regrettable, besides
being contrary the C standard, that some compilers, such as
\texttt{icc}, choose, by default, to assume that NaNs should not be
handled correctly, or that $\pm 0$ can be used interchangeably. We
shall see this on a simple example.

Consider the problem of computing the minimum of two floating-point
numbers. This can be implemented in four straightforward ways:
\begin{enumerate}
\item \verb@x < y ? x : y@
\item \verb@x <= y ? x : y@
\item \verb@x > y ? y : x@
\item \verb@x >= y ? y : x@
\end{enumerate}
These four expressions are equivalent over the real numbers, but they
are not if one takes NaNs into account, or one differentiates $\pm 0$.
Witness:

\begin{center}
\begin{tabular}{|r|r|r|r|r|r|}
\hline
\texttt{x} & \texttt{y} &
  \texttt{x<y ? x:y} &
  \texttt{x<=y ? x:y} &
  \texttt{x>y ? y:x} &
  \texttt{x>=y ? y:x} \\
\hline
+0 & -0 & -0 & 0 & 0 & -0 \\
NaN & 1 & 1 & 1 & NaN & NaN \\
\hline
\end{tabular}
\end{center}

On SSE (see~~\ref{part:SSE}),
both \texttt{icc}, by default, and \texttt{gcc} with the
\texttt{-O2 -ffast-math} option will compile all four expressions as
though they were the first one. The reason is
that the first expression maps directly to one assembly instruction,
\texttt{minss} or \texttt{minsd}, while the others entail more complex
and slower code. Interestingly, if the four above expression are
present in the same function, \texttt{gcc -O2 -ffast-math} will detect
that they are equivalent and simply reuse the same result.

We echo William Kahan%
\footnote{See for instance the essay \emph{The baleful influence of SPEC
    benchmarks upon floating-point arithmetic} on Kahan's web page.}
in deploring that some compilers allow themselves, \emph{by default},
to ignore language standards and apply unsafe optimisations,
presumably in order to
increase performance in benchmarks.
Some algorithms are written in a certain way so as to minimise
roundoff error, and compilers should not
rewrite them in another way. Also, static analysis results obtained from
the source code (see Section~\ref{part:analysers}) may not be
applicable on the object code if the compilers make such unsafe
transformations. We thus suggest that users of analysis tool operating
on the source code read the documentation of their compiler carefully
in search of discussion of ``relaxed'' or ``strict'' floating-point issues.

\subsection{Input/output issues}\label{part:input-output}
Another possible compilation issue is how compilers interpret
constants in source code. The C norm states:
\begin{norm}
For decimal floating constants, and also for hexadecimal floating
constants when \texttt{FLT\_RADIX}%
\footnote{\texttt{FLT\_RADIX} is the radix of floating-point
  computations, thus 2 on IEEE-754 systems. There currently exist few
  systems with other radices.}
is not a power of 2, the result is either
the nearest representable value, or the larger or smaller
representable value immediately adjacent to the nearest representable
value, chosen in an implementation-defined manner. For hexadecimal
floating constants when \texttt{FLT\_RADIX} is a power of 2, the result is
correctly rounded.
\end{norm}
This means that two compilers on the same platform may well interpret the same
floating-point decimal literal in the source code as different floating-point
value, even if both compilers follow C99 closely.
Similar limitations apply to the behaviour
of the C library when converting from decimal representations to
floating-point variables \cite[\S 7.20.1.3]{C99}.

Reading and printing floating-point numbers
accurately is a non-trivial issue if the printing base (here, 10) is
not a power of the computation base (binary in
the case of IEEE-754) \cite{Clinger90,SteeleWhite90}.
There exist few guarantees as to the precision of
results printed in decimal in the C~norm \cite[\S 7.19.6.1, F.5]{C99}.
IEEE-754, however, mandates some guarantees \cite[\S
5.6]{IEC-60559,IEEE-754}, such that printing and reading back the
values should yield the same numbers, within certain bounds.
However, we have seen that the standard C libraries of certain systems
are somewhat unreliable; thus, one may prefer not to trust them on
accuracy issues. Printing out exact values and reading them is
important for replaying exact test cases.

In order to alleviate this, we suggest the use of hexadecimal
floating-point constants, which are interpreted exactly.
Unfortunately, many older compilers do not support these; also, in
order to print floating-point values as hexadecimal easily,
\texttt{printf} and associated functions have to support
the \texttt{\%a} and \texttt{\%A} formats,
which is not yet the case of all current C libraries.

\section{Example}\label{part:example}
Arguably, many of the examples we gave in the preceding sections,
though correct, are somewhat contrived: they discuss small
discrepancies, often happening with very large or very small
inputs. In this section, we give a complete and realistic example of
semantic problems related to differences between floating-point implementations
(even, dependent on compilation options!). It
consists of two parts:
\begin{enumerate}
\item an algorithm for computing a modulo (such as mapping an angle
  into $[-180,180]$ degrees), inspired by an algorithm found in an
  embedded system;
\item possible implementations of tabulated functions.
\end{enumerate}
The composition of the two give seemingly innocuous implementations of angular
periodic functions... which crash for certain specific values of the
inputs on certain platforms.

Given $x$, $m$ and $M$ ($m < M$), we want to compute $r$ such that $r-x$ is an
integer multiple of $M-m$ and $m \leq r \leq M$. The following
algorithm, adapted from code found in a real-life critical system, is
correct if implemented over the real numbers:
\begin{verbatim}
double modulo(double x, double mini, double maxi) {
  double delta = maxi-mini;
  double decl = x-mini;
  double q = decl/delta;
  return x - floor(q)*delta;
}
\end{verbatim}

Let us apply this algorithm to the following case:
\begin{verbatim}
int main() {
  double m = 180.;
  double r = modulo(nextafter(m, 0.), -m, m);
}
\end{verbatim}
We recall that \texttt{floor(\textit{x})} is the greatest integer less
than or equal to \texttt{\textit{x}}, and that \texttt{nextafter(\textit{a},
\textit{b})} is the next representable \texttt{double} value from
\texttt{\textit{a}} in the direction
of~\texttt{\textit{b}}.
\texttt{nextafter(m, 0.)} is thus the greatest double-precision number
strictly smaller than~180.

The above program, compiled by \texttt{gcc} 3.3 with optimisation
for the x87 target, yields an output $r \simeq 179.99999999999997158$.
In such a mode, variable \texttt{q} is cached in a extended precision
register; $\texttt{q} = 1 - \varepsilon$ with $\varepsilon \simeq 7.893.10^{-17}$.
However, if the program is compiled without optimisation, \texttt{decl}
is saved to, then reloaded from, a double-precision temporary variable; in
the process, \texttt{decl} is rounded to $360$, then \texttt{q} is $1$.
The program then returns $r \simeq
-180.00000000000002842$, which is outside the specified bounds.

Simply rewriting the code as follows makes the problem
disappear (because the compiler holds the value in a register):
\begin{verbatim}
double modulo(double x, double mini, double maxi) {
  double delta = maxi-mini;
  return x - floor((x-mini)/delta)*delta;
}
\end{verbatim}
This is especially troubling since this code and the previous one look
semantically equivalent.

The same phenomenon occurs, even if optimisation is turned on,
if we add a logging statement (\texttt{printf()}) after the
computation of \texttt{decl}. This is due to
the forced spilling of the floating-point registers into main memory
across function calls.

Interestingly enough, we discovered the above bug after {\astree}
would not validate the first code fragment with the post-condition
that the output should be between \texttt{mini} and \texttt{maxi}:
{\astree} was giving an interval with a lower bound slightly below
\texttt{mini}. After vainly trying to prove that the code fragment
worked, the author began to specifically search for counter-examples.
In comparison, simply performing unit testing
on a IA32 PC will not discover the problem if the compiler implements
even simple register scheduling (which will be turned on by any
optimisation option). Guidelines for testing generally specify that
programs should be tested on the target system; however,
unit testing on the target architecture will discover this problem only
if using carefully chosen values. The above phenomenon was missed
because testers, following testing guidelines, had tested the program
at points of discontinuity, and \emph{slightly} before and after these
points, but had not tested the values just before and after these points.

Now, one could argue that the odds of landing exactly on
\texttt{nextafter(180., 0.)} are very rare. Assuming an embedded
control routine executed at 100~kHz, 24 hours a day, and a uniform
distribution of values in the $[-180,180]$ interval, such a value
should happen once every $4000$ years or so on a single unit.%
\footnote{For $180$, the unit at the last position is
  $\delta=2^{-45}$; all reals in $(180-(3/2)\delta,180-\delta/2)$ are
  rounded to $180-\delta$, thus the probability of rounding a random
  real in $[-180,180]$ to this number is $360/\delta$.}
However, in the case of a mass-produced system (an automobile model, for
instance), this argument does not hold.
If hundreds of thousands of systems featuring a defective component are
manufactured every year, there will be real failures happening
when the systems are deployed --- and they will be very difficult to
recreate and diagnose.
If the system is implemented in single precision, the odds are
considerably higher with the same probabilistic assumptions: a single
100~Hz system would break down twice a day.

Now, it seems that a slight error such as this should be of no
consequence: the return value is less than the intended lower bound,
but still extremely close to it.
However, let us consider a typical application of computing
modulos: computing some kind of periodic function, for instance
depending on an angle. Such a function is likely to contain
trigonometric operators, or other operations that are
costly and complex to compute. A common way to work around this
cost and complexity is to look such functions up from a precomputed
table.

One implementation technique sometimes found is to put all these
tables in some big array of constants
(perhaps loaded from a file during initialisation), and
read from the array at various offsets: 
\begin{verbatim}
val = bigTable[r + 180 + FUNCTION_F_OFFSET];
\end{verbatim}
This means an implicit truncation of
\verb@r+180+FUNCTION_F_OFFSET@
to 0; if $r$ is a little below $-180$, then
this truncation will evaluate to $\texttt{FUNCTION\_F\_OFFSET}-1$. The
table look-up can then yield whatever value is at this point in memory,
possibly totally out of the expected range.

Another example is a table look-up with interpolation:
\begin{verbatim}
double periodicFunction(double r) {
  double biased = r+180;
  double low = floor(biased);
  double delta = biased-low;
  int index = (int) low;
  return ( table[index]*(1-delta)
         + table[index+1]*delta );
}
\end{verbatim}
If $r$ is slightly below $-180$, the value return will depend on
\texttt{table[-1]}, that is, whatever is in memory before
\texttt{table}. This can result in a segmentation fault (access to a
unallocated memory location), or, more annoyingly, in reading whatever
is at that location, including special values such as \textit{NaN} or
$\pm\infty$, or even simply very large values. With
$\texttt{table[-1]} = 10^{308}$ and $\texttt{table[0]}=0$, the above
program outputs approximately $2.8\times10^{294}$ --- a value
probably too large for many algorithms to handle gracefully.

Such kinds of problems are extremely
difficult to reproduce if they are found in practise ---
they may result in program crashes or
nondeterministic behaviour for very rare input values. Furthermore,
they are not likely to be elicited by random testing.%
\footnote{The example described here actually demonstrates the
  importance of \emph{non-random} testing: that is, trying values that
  ``look like they might cause problems''; that is, typically, values
  at or close to discontinuities in the mathematical function
  implemented, special values, etc.}
Finally, the problem may disappear if the program is tested with
another compiler, compilation options, or execution platform.

\section{A few remarks on Java}\label{part:Java}
Java's early floating-point model was a strict subset of
  IEEE-754 \cite[\S 4.2.3]{JavaSpec1}: essentially, strict IEEE-754
single and double-precision arithmetic without the
 exception traps (overflow, invalid operation\dots)
and without rounding modes other than round-to-nearest. 
However, strict compatibility with
IEEE-754 single- and double-precision operations is difficult to
achieve on x87 (see Section~\ref{part:x87}).
As a consequence, requests were
made so that strict compatibility would be relaxed in order to get
better performance, particularly for scientific computing
applications. The possibility of giving up Java's deterministic,
portable semantics was requested by some \cite{KahanDarcy98},
but controversial for others \cite{JavaGrandeSC98}.
Finally, the Java language specification was altered
\cite[\S 4.2.3]{JavaSpec2}: run-time computing in extra precision
(single-extended and double-extended formats) was allowed for classes
and methods not carrying the new \texttt{strictfp} modifier \cite[\S
15.4]{JavaSpec2,Java_language_spec3}: they may be evaluated with an extended exponent range. To summarise, the Java designers made it legal to compile expressions to straightforward x87~code. 

This is actually a bolder decision than may appear at first sight. The whole design of the Java language, with well-defined semantics, aimed at enforcing unique semantics for single-threaded programs%
\footnote{Java does not attempt to prescribe how a multi-threaded
  program should be executed: it describes valid multi-threading
  behaviours, all of which are equally
  acceptable~\cite[Chapter~17]{Java_language_spec3}. Some
  nondeterminism is thus acceptable.}
that do not use system-dependent features (files, graphical interfaces, etc.): a program that merely did single-threaded computations should behave identically regardless of the machine, operating system and Java runtime system. This decision breaks this uniqueness of semantics and introduces platform dependencies, which Java was supposed to overcome.

Let us take for instance the \texttt{foo} program of
Section~\ref{part:x87}, translated into Java:
\begin{verbatim}
class Foo {
  static double foo(double v) {
    double y = v*v;
    return (y / v);
  }

  public static void main(String[] args) {
    System.out.println(foo(1E308));
  }
}
\end{verbatim}

We use \texttt{gcj} (the Java compiler associated with \texttt{gcc})
version~4.1.1 on a x86\_64 system.
Unsurprisingly, when compiled to machine code
for the SSE target, the program prints \texttt{Infinity}.
The results get more diverse on the x87 target:
\begin{itemize}
\item If no optimisation is used, the program prints \texttt{Infinity}. Examination of the assembly code shows that intermediate values are spilled to memory.
\item If optimisation level~1 is used (\texttt{-O}), then the program prints \texttt{1E308} ($10^{308}$). Intermediate values are left in registers, but functions are not inlined.
\item If optimisation level~3 is used (\texttt{-O3}), the program prints \texttt{Infinity}. The compiler inlines function \texttt{foo} and detects the parameter to \texttt{System.out.println()} is a constant. It computes that constant using strict IEEE-754 double precision, thus the result.
\end{itemize}

The \texttt{strictfp} modifier should force the program to adopt strict IEEE-754 round-to-nearest semantics. However, this modifier is ignored by \texttt{gcj}~\cite[\S 2.1]{gcj} (and the above experiments show identical results regardless of \texttt{strictfp}). The author is even unsure that the behaviour noted above is correct even in the absence of \texttt{strictfp}: \texttt{strictfp} is supposed to affect only temporaries~\cite[\S 15.4]{Java_language_spec3}, and \texttt{y} is not a temporary.

The ``normal'' mode of operation of Java is not straight compilation to native code, but compilation to bytecode; the bytecode may then be executed inside a Java Runtime Environment (JRE), comprising a Java Virtual Machine (JVM). The simplest JVMs just interpret the bytecode, but more advanced ones do ``just-in-time compilation'' (JIT). A JIT-capable virtual machine will typically interpret bytecode at first, but will detect frequently used functions and will compile these to native codes. Because JIT occurs at runtime, it can perform advanced optimisations, such as detecting that certain arguments occur frequently for certain functions and specialising these functions, that is, compiling a special version of these functions for certain values of its arguments and replacing the calls to the generic function by calls to the specialised function when appropriate.
The same kind of problems that we gave above for generation of native code can thus occur: the interpreted version will spill temporaries to memory and output one value, a JIT-compiled version may give another value, but if JIT specialises the function it may give a third value (or the same value as if interpreted). Note that JIT can thus dynamically change the semantics of a function for reasons unrelated to the program being executed: the programmer has no means to predict or control when and how JIT compilation is performed.

We can conclude that programmers should be cautious before assuming that Java programs will behave predictably when using floating-point. First, this is only true if the \texttt{strictfp} modifier is used, and, even then, some very popular compilers and runtime systems (\texttt{gcj} ships by default with many GNU/Linux systems) ignore this modifier and may even ignore some other parts of the specification. The presence of JIT compilers may also add various amusing effects.

\section{Implications for program verification}\label{part:implications}
Program verification comprises a variety of methods whose goals is to
prove formally that programs fit their specifications, often lumped
together under the term ``formal methods''. Formal methods have long
ignored floating-point computations, because they were judged too
baroque or too difficult to model.

\subsection{Goals of program verifications}
Purposes of verification of programs using floating-point computations
may be, in increasing order of complexity:
\begin{enumerate}
\item Proving that the program will never trigger ``undefined'' or
  ``undesirable'' behaviours, such as an overflow on a conversion from
  a floating-point type to an integer type.
  This problem has attracted the attention of both the industrial
  world and the program analysis community since the much publicised
  self-destruction of the Ariane~5 launcher during its maiden flight,
  which was due to overflow in a conversion from a 64-bit floating
  point value to a 16-bit integer~\cite{Flight_501_report}.

  Proving that a value does not overflow entails finding some bounds
  for that value, and if that value is the result of a complex
  computation depending on the history of previous inputs and outputs
  (as is the case of e.g. infinite impulse response filters, rate
  limiters, and combinations thereof), then finding and proving such bounds
  entails proving the stability of the numerical computation.
  In many cases, though, automatic \emph{program analysers}, taking as
  input the source code of the program (typically in~C or some other
  language, perhaps even in assembly language) can automatically compute
  properties of such systems. If the analyser is  \emph{sound}, then all the
  properties it prints (say, $x < 5$)
  hold for every run of the analysed programs.
  
  The {\astree} system
   analyses programs
  written in a subset of the
  C~programming language
  \cite{ASTREE_ESOP05,BlanchetCousotEtAl_PLDI03,BlanchetCousotEtAl02-NJ}
  and attempts bounding all variables and
  proving the absence of overflows and other runtime errors. While it
  is possible to specify assertions (such as bounds on some variables
  representing physical quantities with known limits),
  which {\astree} will attempt to prove, the system is not designed to
  prove such user-defined properties.

\item Pin-pointing the sources of roundoff errors in the program;
  proving an upper bound on the amount of roundoff error in some
  variable.

  While in the preceding class of problems, it does not matter whether
  the numerical computation makes sense as long as it does not crash
  or violate some user-specified assertion, here the problems are
  subtler. The \textsc{Fluctuat} tool automatically
  provides results on the origin
  and magnitude of roundoff errors in numerical
  programs~\cite{Martel_HOSC2006,Goubault_SAS01,Martel_SAS02,Martel_ESOP02}.

\item Proving that the program implements such or such numerical
  computation up to some specified error bound. Except in the simplest
  cases, automated methods are unsuitable; if formal proofs are
  desired, then computerised proof assistants may help. Proving that a
  program fits its specification
  is difficult in general, even more so over numerical programs, which
  do not have ``good'' algebraic properties;
  yet some progress has been recently made in that
  direction~\cite{Filliatre_Boldo_ARITH18}.
\end{enumerate}
 
\subsection{Semantic bases of program analysis}
\label{part:semantic-bases}
In order to provide some proofs in the mathematical sense, one has to
start with a mathematical definition of program behaviours, that is, a
\emph{semantics}.~\cite{Winskel_1993}
This semantics should model all possible concrete
behaviours of the system, without omitting any.

An alternative point of view is that, for the sake of simplicity of
verification, the chosen semantics may fail to reflect some behaviours
of the concrete system.
For instance, one may consider
that all floating-point variables behave as real numbers. We have
shown, however, in Sec.~\ref{part:example}, that very real bugs can
occur in simple, straightforward programs that are correct if
implemented over the real numbers, but that exhibit odd, even fatal,
behaviours due to floating-point roundoff. Such an approach is thus
risky if the goal is to provide some assurance that the program
performs correctly, but it may be suitable for \emph{bug-finding}
systems, whose goal is not to prove the absence of bugs, but to direct
programmers to probable bugs. Indeed, bug-findings techniques are
typically \emph{unsound}: they trade soundness (all possible bugs
should be listed) for ease of use (not too many warnings about bugs
that do not exist) and efficiency (quick analysis). In that context,
methods that consider that floating-point variables contain real
numbers (as opposed to floating-point values),
or that integer variables contain unbounded integers (as opposed to
$n$-bit integers computed using modular arithmetic), may be relevant.

In this paper, we are concerned with \emph{sound} verification
techniques: techniques that only produce correct results; that is, if
the analysis of a program results in the analyser claiming that some
conversion from floating-point to integer will not overflow, then it
should not be possible to have this conversion overflow, regardless of
inputs and roundoff errors.
Given the various misunderstandings about floating-point that we cited
in the previous sections, it is no surprise that it is extremely easy
for an analysis designer to build an unsound static analysis tool
without meaning it, for instance by starting with a semantics of
floating-point operations that does not model reality accurately enough.

It is well-known that any method for program verification cannot be at
the same time sound (all results produced are truthful), automatic (no
human intervention), complete (true results can always be proved)
and terminating (always produces a
result) ~\cite{Cousot_methods_logics},%
\footnote{The formal version of this result is a classic of recursion
  theory, known as Rice's theorem:
  let $C$ be a collection of partial recursive functions of one
  variable, then, noting $\phi_x$ the partial recursive function
  numbered~$x$, then $\{x \mid \phi_x \in C\}$ has a recursive
  characteristic function if and only if $C$ is empty or contains all
  partial recursive functions of all
  variables.~\cite[p.~34]{Rogers}\cite[corollary~B]{Rice_1953}
  This result is proved by reduction to the halting problem.}
unless one supposes that
program memory is finite and thus that the system is available to
\emph{model-checking} techniques. Given that the state spaces of
programs with floating-point variables are enormous even with
small numbers of variables, and that the Boolean functions
implementing floating-point computations are not very regular, it
seems that model-checking for whole floating-point
algorithms should not be tractable. As a consequence, we must content
ourselves with techniques that are unsound, non-terminating, incomplete, or not
automatic, or several at the same time. The purpose of this section is
to point out how to avoid introducing unsoundness through carelessness.

All terminating automatic program analysis
methods are bound to be incomplete; that is, they may
be unable to prove certain true facts. Incompleteness is equivalent to
considering a system that has more behaviours than the true concrete
system. Since we must be incomplete anyway, it is as well that we take
this opportunity to simplify the system to make analysis more
tractable; in doing so, we can still be sound as long as we only add
behaviours, and not remove any.
For instance, in {\astree}, most analyses do not attempt to track
exactly (bit-by-bit)
the possible relationships between floating-point values, but
rather rely on the bound on roundoff given by
inequality~\ref{eqn:rounding2}.

\subsection{Difficulties in defining sound semantics}
We have seen many problems regarding the definition of sound
semantics for programs using floating-point; that is, how to attach to
each program a mathematical characterisation of what it actually
does. The following approaches may be used:
\begin{itemize}
\item
A naive approach to the concrete semantics of programs running on
``IEEE-754-compatible'' platforms is to consider that a \texttt{+},
\texttt{-}, \texttt{*} or \texttt{/} sign in the source code, between
operands of type \texttt{float} (resp. \texttt{double}), corresponds to a
strict IEEE-754 $\oplus$, $\ominus$, $\otimes$, $\oslash$ operation, with
single-precision (resp. double-precision) operands and result:
$a \oplus b = r(a + b)$ where $r$ rounds to the target precision.
As we have seen, this does not hold in many common cases, especially
on the x87 (Section~\ref{part:x87}).

\item
A second approach is to analyse assembly or object code, and take the
exact processor-specified semantics as the concrete semantics for each
operation. This is likely to be the best solution if the compiler is
not trusted not to make ``unsafe optimisations''
(see~\S \ref{part:optimisations}). It is possible to
perform static analysis directly on the assembly code, or even object
code~\cite{Gogul_CC04}. In addition, it
is possible, if source code in a high level language is available, to
help the assembly-level static analyser with information obtained from
source-level analysis, through invariant translation
\cite{XR_VMCAI_2003}, which may be helpful for instance for pointer
information.

One remaining difficulty is that some ``advanced'' functions in
floating-point units
may have been defined differently in successive generations of processors, so
we still cannot rule out discrepancies. However,
when doing analysis for embedded systems, the
exact kind of target processor is generally known, so it is possible
to use information as to its behaviour. Another possibility is to
request that programmers do not use poorly-specified functions of the
processor.

\item
A third approach is to encompass all possible semantics of the source
code into the analysis.
Static analysis methods based on abstract interpretations
(Section~\ref{part:analysers}) are
well-suited for absorbing such ``implementation-defined'' behaviours
while still staying sound.
\end{itemize}

\subsection{Hoare logic}
\label{part:hoare}
As an example in the difficulty of proposing simple semantics based on
the source code of the program, let us consider the proof rules of
the popular Hoare logic~\cite{Hoare_1969,Winskel_1993}. These rules
are the basis of many proof assistants and other systems for proving
properties on imperative programs. A Hoare triple
$\HoareTriple{A}{c}{B}$ is read as follows: if the program state at
the beginning of the execution of program $c$ verifies property~$A$,
then, if $c$ terminates, then the final program state verifies
property~$B$. A rule
\begin{equation*}
\infer[r]{C}{H_1 \dots H_n}
\end{equation*}
reads as: if we can prove the hypotheses $H_1,\dots,H_n$, then we can
prove the conclusion $C$ by applying rule~$r$. If there are zero
hypotheses, then we say rule $r$ is an axiom.

We recall the classical rules for assignment, sequence of operation, and test:
\begin{eqnarray*}
\infer[\text{assign}]{\HoareTriple{B[x \mapsto a]}{\assign{x}{a}}{B}}{}
\\
\infer[\text{sequence}]{\HoareTriple{A}{c_0; c_1}{C}}{
  \HoareTriple{A}{c_0}{B} &
  \HoareTriple{B}{c_1}{C}}
\\
\infer[\text{if}]{\HoareTriple{A}{\cifthenelse{b}{c_0}{c_1}}{B}}{
  \HoareTriple{\band{A}{b}}{c_0}{B} &
  \HoareTriple{\band{A}{\bneg{b}}}{c_1}{B}}
\end{eqnarray*}

In the following rule, we use hypotheses of the form
$\vdash C$, reading ``it is possible to prove $C$ in the
underlying mathematical logic''. Indeed, Hoare logic, used to reason
about programs, is parametrised by an underlying mathematical logic,
used to reason about the quantities inside the program.
\begin{equation*}
\infer[\text{weakening}]{\HoareTriple{A}{P}{B}}{\vdash A \implies A' &
  \HoareTriple{A'}{P}{B'} &
  \vdash B' \implies B}
\end{equation*}

These rules \emph{sound} suitable for proving properties of floating-point
programs, if one keeps floating-point expressions such as $x \oplus y$
inside formulae (if multiple floating-point types are used, then one
has to distinguish the various precisions for each operator, e.g.
$x \oplus_d y$ defined as $r_d(x+y)$).

Hoare logic is generally used inside a \emph{proof assistant}, a
program which allows the user to prove properties following the rules,
possibly with some partial automation.
The proof assistant must know how to reason about floating-point quantities.
$x \oplus y$ may be defined as $r(x+y)$, where $r$ is the appropriate
rounding function. $x+y$ is the usual operation over the reals (thus,
the underlying mathematical logic of the Hoare prover should include a
theory of the reals), and $r$ is the rounding function, which may be
defined axiomatically. Inequalities such as ineq.~\ref{eqn:rounding1}, or
properties such as $r \circ r = r$ then appear as lemmas of the
mathematical prover.
An axiomatisation of the floating-point numbers is, for instance, used in the
\textsc{Caduceus} tool~\cite{Filliatre_Boldo_ARITH18}.

However, the above rules, as well as all tools based on them, including
\textsc{Caduceus}, are unsound when applied to programs running
on architectures such as the~x87.
They assume that any given arithmetic or Boolean expression has a
unique value, depending only the value of all variables involved,
and that this value
does not change unless at least one of the variables involved changes
through an assignment (direct or through pointers).
Yet, in the program \verb@zero_nonzero.c@ of Sec.~\ref{part:x87}, we
have shown that it is possible that inequality \texttt{z != 0} holds
at a program line, then, without any instruction assigning anything to
\texttt{z} in whatever way, that inequality ceases to hold. Any proof
assistant straightforwardly based on Hoare logic would have accepted
proving that the assertion in \verb@zero_nonzero.c@ always holds,
whereas it does not in some concrete executions.

Such tools based on ``straightforward'' Hoare rules are thus sound only
on architectures and compilation schemes for which the points where
rounding takes place are precisely known. This excludes:
\begin{itemize}
\item The x87 architecture, because of rounding points depending on
  register scheduling and other circumstances not apparent in the
  source code.
\item Architectures using a fused multiply-add instruction, such as
  the PowerPC (Sec.~\ref{part:PowerPC}), because one does not know
  whether $x \otimes y \oplus z$ will get executed as
  $r(r(x \times y) + z)$ or $r(xy+z)$.
\end{itemize}

What can we propose to make these tools sound even on such architectures?
The first possibility is to have tools operate not on the C source code
(or, more generally, any high-level source code) but on the generated assembly
code, whose semantics is unambiguous.%
\footnote{At least for basic operations $\oplus$, $\ominus$,
  $\otimes$, $\oslash$, $\sqrt{}$. We have seen in
  Section~\ref{part:transcendental} that transcendental
  functions may be less well specified.}
However, directly checking assembly code
is strenuous, since code in assembly language
is longer than in higher-level languages, directly deals with many issues
hidden in higher-level languages (such as stack allocation of variables),
and exhibits many system dependencies.

Another solution for the x87 rounding issue,
is to replace the ``simple'' axiomatic semantics given above
by a nondeterministic semantics where the non-determinacy of the position of
rounding is made explicit. Instead of an assignment $\assign{x}{a}$
being treated as a single step, it is broken into a sequence of
elementary operations. Thus, $\assign{x}{a \oplus (b \otimes c)}$,
over the double precision numbers, is
decomposed into
$\assign{t}{b \otimes c}; \assign{x}{a \oplus t}$.
Then, each operation is written using the $r_e$ extended
precision rounding function and the $r_d$ double precision rounding
function, taking into account the nondeterminism involved,
noting $\ndt$ a function returning \textsf{true} or
\textsf{false} nondeterministically and $\nop$ the instruction doing
nothing:
\begin{center}
\begin{minipage}{10cm}
$\assign{t}{r_e(b \times c)}$;\\
$\cifthenelse{\ndt}{\assign{t}{r_d(t)}}{\nop}$;\\
$\assign{x}{r_e(a + t)}$
\end{minipage}
\end{center}

The transformation consists in using the $r_e$ rounding
operation in elementary operations (thus $a \oplus b$ gets translated
as $r_e(a+b)$) and prepending an optional (nondeterministically
chosen) $r_d$ rounding step before any operation on any operand on
that operation. The resulting code is then suitable for proofs in
Hoare logic, and the properties proved are necessarily fulfilled by
any compilation of the source code, because, through nondeterminism,
we have encompassed all possible ways of compiling the operations.
We have thus made the proof method sound again, at the expense of
introduced nondeterminism (which, in practise, will entail
increased complexity and tediousness when proving properties) and
also, possibly,
of increased incompleteness with respect to a particular target (we
consider behaviours that cannot occur on that target, due to the way
the compiler allocates registers or schedules instructions).

If we know more about the compiler or the application binary
interface, we can provide more precise semantics (less
nondeterminism, encompassing fewer cases that cannot occur in
practice). For instance, using
the standard parameter passing scheme on IA32 Linux
for a non-inline function,
floating-point values are passed on the stack. Thus, a Hoare logic
semantics for parameter passing would include a necessary $r_s$
(single precision) or $r_d$ rounding phase, which could simplify the
reasoning down the program, for instance by using the fact that
$r_s \circ r_s = r_s$ and $r_d \circ r_d = r_d$.

A similar technique may be used to properly handle compound
instructions such as fused-multiply-add. On a processor where
fused-multiply-add yields $r(a \times b + c)$, there are two ways of
compiling the expression $(a \otimes b) \oplus c$: as
$r(r(a \times b) + c)$ or as $r(a \times b + c)$. 
By compiling floating-point expressions (using $\oplus$, $\otimes$ etc.)
into a nondeterministic choice between unambiguous expressions over
the reals (using $r$, $+$, $\times$ etc.) we get a program amenable to
sound Hoare proof techniques. Again, the added nondeterminism is
likely to complicate proofs.

This technique, however, relies on the knowledge of how the compiler
may possibly group expressions. Because compilers can optimise code
across instructions and even across function calls, it is likely that
the set of possible ways of compiling a certain code on a certain
platform is large.

To summarise our findings:
the use of Hoare-logic provers is hampered by platforms
or language where a given floating-point expression does not have a
single, unambiguous meaning; straightforward application of the rules
may yield unsound results, and workarounds are possible, but costly.

\subsection{Static analysers based on abstract interpretation}
\label{part:analysers}
We here consider methods based on \emph{abstract interpretation}, a generic
framework for giving an \emph{over-approximation} of the set of
possible program executions.~\cite{CousotCousot92} An \emph{abstract
domain} is a set of possible symbolic constraints with which to
analyse programs. For instance, the interval domain represents
constraints of the form $C_1 \leq x \leq C_2$,
the \emph{octagon abstract domain}~\cite{Mine_AST01}
constraints of the form
$\pm x \pm y \leq C$ where $x$ and $y$ are program variables and the
$C$ are numbers.

\subsubsection{Intervals}
The most basic domain for the analysis of floating-point programs is
the interval domain
\cite{BlanchetCousotEtAl02-NJ,Mine_PhD,Mine_ESOP04}: to each
quantity $x$ in the program, attach an interval $[m_x,M_x]$ such that
in any concrete execution, $x \in [m_x,M_x]$. Such intervals
$[m_x,M_x]$ can be computed efficiently in a static analyser for
software running on a IEEE-754 machine if one
runs the analyser on a IEEE-754 machine, by doing interval arithmetic
on the same data types as used in the concrete system, or smaller ones.

Fixing $x$, $y \mapsto x \oplus y$ is
monotonic. It is therefore tempting, when implementing interval
arithmetic, to approximate $[a,b] \oplus [a',b']$ by
$[a \oplus a', b \oplus b']$ with the same rounding mode.
Unfortunately, this is not advisable unless one is really sure that
computations in the program to be analysed are really done with the
intended precision (no extended precision temporary values, as common
on x87, see Sec.~\ref{part:x87}), do not suffer from double rounding effects
(see~\ref{part:double-rounding} and the preceding section),
and do not use compound operations
instead of atomic elementary arithmetic operations. In contrast, it is
safe to use directed rounding:
computing upper bounds in round-to-$+\infty$ mode, and lower bounds
in round-to-$-\infty$ mode:
$[a,b] \oplus [a',b'] = [a \oplus_{-\infty} a', b \oplus_{+\infty} b']$.%
\footnote{Changing the rounding mode may entail significant efficiency
  penalties. A common trick is to set the processor in
  round-to-$+\infty$ mode permanently, and replace $x \oplus_{-\infty} y$ by
$-((-x) \oplus_{+\infty} (-y))$ and so on.}
This is what is done in {\astree} \cite{ASTREE_ESOP05}. Note, however,
that on some systems, rounding-modes other than round-to-nearest are
poorly tested and may fail to work properly (see
Section~\ref{part:buggy_rounding_modes}); the implementers of analysers
should therefore pay extra caution in that respect.

Another area where one should exercise caution is strict
comparisons. A simple solution for abstracting the outcome for
\texttt{x} of \verb+x < y+, where $\texttt{x} \in [m_x,M_x]$
and $\texttt{y} \in [m_y,M_y]$, is to take $M'_x = \min(M_x,M_y)$, as
if one abstracted \verb+x <= y+. Unfortunately, this is not sufficient
to deal with programs that use special floating-point values as
markers (say, $0$ is a special value meaning ``end of table''). One
can thus take $M'_x = \min(M_x,\textit{pred}(M_y))$ where
$\textit{pred}(x)$ is the largest floating-point number less than
$x$ \emph{in the exact floating-point type used}. However, as we
showed in \S\ref{part:x87}, on x87 a comparison may be performed on
extended precision operands, which may later be rounded to lesser
precisions depending on register spills. We showed in the preceding
section that this particularity made some proof methods unsound, and
it also makes the above bound unsound.

In order to be sound, we
should use the \textit{pred} operation on the extended precision type;
but then, because of possible rounding to lesser precisions, we should
round the bounds of the interval... which would bring us back to
abstracting $<$ as we abstract~$\leq$.

We can, however, refine the
technique even more. If we know that \texttt{x} is exactly
representable in single (resp. double) precision, then we can safely
use the \textit{pred} function on single (resp. double) precision floats.
Some knowledge of the compiler and the application binary interface may
help an analyser establish that a variable is exactly representable in
a floating-point type; for instance, if a function is not inlined and
receives a floating-point value in a \texttt{double} variable passed
by value on the stack, then this value is exactly representable in double
precision. More generally, values read from memory without the
possibility that the compiler has ``cached'' them inside a register
will necessarily be representable in the floating-point format
associated with the type of the variable.

\subsubsection{Numerical relational domains}
The interval domain, while simple to implement, suffers from not
keeping relations between variables. For many applications, one should
also use relational abstract domains --- abstract domains capable of
reflecting relations between two or more variables.
Relational domains however tend to be designed for data taken
in ideal structures (ordered rings,
etc.); it is thus necessary to bridge the gap between these ideal
abstract structures and the concrete execution. Furthermore, it is necessary
to have effective, implementable algorithms for the abstract domains.

Min\'e proposed three abstraction steps~\cite{Mine_PhD,Mine_ESOP04}:
\begin{enumerate}
\item From a (deterministic) semantics over floating-point numbers to
  a nondeterministic semantics over the real numbers, taking into
  account the rounding errors.
\item From the ``concrete'' nondeterministic semantics over the
  reals, to some ideal abstract domain over the reals ($\bbR$) or the
  rationals ($\bbQ$), such as the
  octagon abstract domain \cite{Mine_AST01}.
\item Optionally,
  the ideal abstract domain is over-approximated by some effective
  implementation. For instance, wherever a real number is computed in
  the ideal domain, a lower or upper bound, or both, is computed in
  the implementation, for instance using floating-point arithmetic
  with directed rounding.
\end{enumerate}

Following a general method in abstract interpretation
\cite{Cousot97-1}, the succession of abstraction steps yields a
hierarchy of semantics:
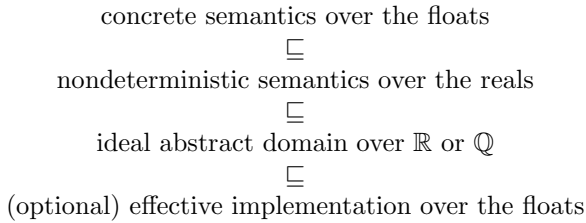
\begin{center}
concrete semantics over the floats\\
$\sqsubseteq$\\
nondeterministic semantics over the reals\\
$\sqsubseteq$\\
ideal abstract domain over $\bbR$ or $\bbQ$\\
$\sqsubseteq$\\
(optional) effective implementation over the floats
\end{center}
The analysis method is sound (no program behaviour is ignored),
but, because of the abstractions, is incomplete (the analyser takes into account
program behaviours that cannot appear in reality, thus is incapable of
proving certain tight properties).

The first step is what this paper is concerned about: a sound modelling of
the floating-point computations. If this step is not done is a sound manner,
for instance if the semantics of the programming language or the target
platform with respect to floating point are misunderstood, then the whole
analysis may be unsound.

The simplest solution, as proposed by Min\'e and implemented
in {\astree}, is to consider that the ``real''
execution and the floating-point execution differ by a small
error, which can be bound as recalled in~\ref{part:rounding}
($|x-r(x)| \leq \erel.|x| + \eabs$).
However, in doing so, one must be careful not to be too optimistic:
for instance, because of possible double rounding,
the error bounds suitable for simply rounded round-to-nearest
computations may be incorrect if, due to possible ``spills'', double
rounding may occur (see~\ref{part:double-rounding}). The $\erel$ coefficient
must thus be adjusted to compensate possible double rounding.

The second step depends on the abstract domain.
For instance, for the octagon abstract domain \cite{Mine_AST01}, it
consists in a modified shortest-path algorithm. In any case, it should
verify the soundness property: if a constraint (such as $x + y < 3$)
is computed, then this constraint should be true on any concrete
state. Finally, the last step is usually achieved using directed
rounding.

\subsubsection{Symbolic abstract domains}
Other kinds of relational domains propagate pure constraints or
guards (say, $\texttt{x} = \texttt{y} \oplus \texttt{z}$). An
example is Min\'e's symbolic rewrite domain implemented in
{\astree}~\cite{Mine_PhD}.
Intuitively, if some arithmetic condition is true on some
variables $x$ and $y$, and neither $x$ nor $y$ are suppressed, then
the condition continues to hold later. However, we have seen in
Section~\ref{part:IA32} that one should beware of ``hidden'' rounding
operations, which may render certain propagated identities invalid.
We may work around this issue using the same techniques as
in Sec.~\ref{part:hoare}: allow rounding at intermediate program points.

In addition, we have also seen that certain equivalences such as
$x \ominus y = 0 \iff x = y$ are not necessarily valid, depending on
whether certain modes such as flush-to-zero are active.%
\footnote{There is still the possibility of considering that
$x \ominus y = 0 \implies |x - y| < 2^{E_{\min}}$.}

\subsection{Testing}
In the case of embedded system development,
the development platform (typically, a PC running some
variant of IA32 or x86\_64 processors) is not the same as the target
platform (typically, a microcontroller). Often, the target platform is
much slower, thus making extensive unit testing time-consuming;
or there may be a limited number of them for the
development group. As a consequence, it is tempting to test or debug
numerical programs on the development platform. We have shown that this
can be a risky approach, even if both the development platform and the
target platform are ``IEEE-754 compatible'', because the same program
can behave differently on two IEEE-754 compatible platforms.

We have also seen that in
some cases, even if the testing and the target platform are
identical, the final result may depend on the vagaries of
compilation.
One should be particularly cautious on platforms,
such as IA32 processors with the x87
floating-point unit, where the results of computations can
depend on register scheduling. 
Even inserting ``monitoring'' instructions can affect the
final result, because these can change register allocation, which
can change the results of computations on the x87. This is especially
an issue since it is common to have ``debugging-only'' statements
excluded from the final version loaded in the system.
In the case of the x87, it is possible
to limit the discrepancies (but not totally eradicate them) by setting the
floating-point unit to 53-bit mantissa precision, if one computes
solely with IEEE-754 double precision numbers.

Static analysis techniques where concrete traces are replayed inside
the analysis face similar problems. One has to have an exact semantic
model of the program to be analysed as it runs on the target platform.

\section{Conclusion}
Despite claims of conformance to standards, common
software/hardware platforms may exhibit subtle differences with respect to
floating-point computations. These differences pose special problems
for unit testing or debugging, unless one uses exactly the same object
code as the one executed in the target environment.

More subtly, on some platforms, the exact same expression, with the
same values in the same variables, and the same compiler,
can be evaluated to different results, depending on seemingly
irrelevant statements (printing debugging information or other
constructs that do not openly change the values of variables). This is
in particular the case on the Intel 32-bit platform,
due to the way that temporary variables are commonly handled by
compilers. This breaks an assumption, common in program verification
techniques, that, barring the modification of some of the variables
used in the expression (directly or through a pointer), the same
expression or test evaluates twice to the same value. We have seen
that this assumption is false, even providing concrete example where a
condition holds, then ceases to hold two program lines later, with no
instruction in the source code that should change this condition in between.

We proposed some alterations to program verification techniques based
on Hoare logic in order to cope with
systems where this assumption does not hold. However, these
alterations, necessary in order to preclude proving false statements
on programs, make proofs more complex due to increase nondeterminism.
With respect to automated analysis techniques based on abstract
interpretation, we explained how to cope with the Intel 32-bit platform
in a sound way, including with precise analysis of strict
comparisons, through simple techniques.

With respect to the development of future systems, both hardware, and
software, including compilers, we wish to stress that reproducibility
of floating-point computations is of paramount importance for
safety-critical applications.
All designs --- such as the common compilation schemes on the Intel 32-bit
platform --- where the results can change depending on seemingly irrelevant
circumstances such as the insertion of a ``print'' statement make
debugging fringe conditions in floating-point programs difficult,
because the insertion of logging instructions may
perturb results. They also make several program analysis techniques
unsound, leading to sometimes expensive fixes.

The fact that some well-known compilers, by default, decide to
activate optimisations that contradict published programming language
standards and also break verification techniques, is not
comforting. ``Unsafe'' optimisations often yield improved performance,
but their choice should be conscious.

\section*{Acknowledgements}
We thank Patrick Cousot, Pascal Cuoq, J\'er\^ome Feret, David Madore, Antoine
Min\'e and the anonymous referees for their helpful comments.

\bibliography{floating-point}

\begin{thebibliography}{49}
\expandafter\ifx\csname natexlab\endcsname\relax\def\natexlab#1{#1}\fi
\expandafter\ifx\csname url\endcsname\relax
  \def\url#1{{\tt #1}}\fi

\bibitem[Adv(2005)]{AMD64-1}
{\em {AMD64} Architecture Programmer's Manual Volume~1: Application
  Programming}.
\newblock Advanced Micro Devices, 3.10 edition, March 2005.

\bibitem[Appel and Ginsburg(1997)]{Appel_modern_compiler_C_97}
Andrew Appel and Maia Ginsburg.
\newblock {\em Modern compiler implementation in {C}}.
\newblock Cambridge University Press, revised and expanded edition, December
  1997.
\newblock ISBN 052158390X.

\bibitem[Balakrishnan and Reps(2004)]{Gogul_CC04}
Gogul Balakrishnan and Thomas~W. Reps.
\newblock Analyzing memory accesses in x86 executables.
\newblock In {\em CC}, volume 2985 of {\em LNCS}. Springer, 2004.
\newblock ISBN 3-540-21297-3.

\bibitem[Blanchet et~al.(2002)Blanchet, Cousot, Cousot, Feret, Mauborgne,
  Min{\'e}, Monniaux, and Rival]{BlanchetCousotEtAl02-NJ}
B{.} Blanchet, P{.} Cousot, R{.} Cousot, J{.} Feret, L{.} Mauborgne, A{.}
  Min{\'e}, D{.} Monniaux, and X{.} Rival.
\newblock Design and implementation of a special-purpose static program
  analyzer for safety-critical real-time embedded software.
\newblock In {\em The Essence of Computation: Complexity, Analysis,
  Transformation}, number 2566 in LNCS, pages 85--108. Springer, 2002.

\bibitem[Blanchet et~al.(2003)Blanchet, Cousot, Cousot, Feret, Mauborgne,
  Min{\'e}, Monniaux, and Rival]{BlanchetCousotEtAl_PLDI03}
B{.} Blanchet, P{.} Cousot, R{.} Cousot, J{.} Feret, L{.} Mauborgne, A{.}
  Min{\'e}, D{.} Monniaux, and X{.} Rival.
\newblock A static analyzer for large safety-critical software.
\newblock In {\em PLDI}, pages 196--207. ACM, 2003.

\bibitem[Caspi et~al.(1987)Caspi, Pilaud, Halbwachs, and Plaice]{LUSTRE}
Paul Caspi, Daniel Pilaud, Nicolas Halbwachs, and John~A. Plaice.
\newblock {LUSTRE}: a declarative language for real-time programming.
\newblock In {\em POPL '87: Proceedings of the 14th ACM SIGACT-SIGPLAN
  symposium on Principles of programming languages}, pages 178--188. ACM Press,
  1987.
\newblock ISBN 0-89791-215-2.

\bibitem[Clinger(1990)]{Clinger90}
William~D. Clinger.
\newblock How to read floating point numbers accurately.
\newblock In {\em PLDI}, pages 92--101. ACM, 1990.
\newblock ISBN 0-89791-364-7.

\bibitem[Cormen et~al.(1990)Cormen, Leiserson, and Rivest]{Cormen}
Thomas~H. Cormen, Charles~E. Leiserson, and Ronald~L. Rivest.
\newblock {\em Introduction to algorithms}.
\newblock MIT Press, 1990.

\bibitem[Cousot(1997)]{Cousot97-1}
P{.} Cousot.
\newblock Constructive design of a hierarchy of semantics of a transition
  system by abstract interpretation.
\newblock {\em ENTCS}, 6, 1997.

\bibitem[Cousot(1990)]{Cousot_methods_logics}
Patrick Cousot.
\newblock Methods and logics for proving programs.
\newblock In Jan van Leeuwen, editor, {\em Handbook of theoretical computer
  science}, volume~B, chapter~15, pages 841--994. MIT Press, 1990.

\bibitem[Cousot and Cousot(1992)]{CousotCousot92}
Patrick Cousot and Radhia Cousot.
\newblock Abstract interpretation and application to logic programs.
\newblock {\em J. Logic Prog.}, 2-3\penalty0 (13):\penalty0 103--179, 1992.

\bibitem[Cousot et~al.(2005)Cousot, Cousot, Feret, Mauborgne, Min\'e, Monniaux,
  and Rival]{ASTREE_ESOP05}
Patrick Cousot, Radhia Cousot, J\'er\^ome Feret, Laurent Mauborgne, Antoine
  Min\'e, David Monniaux, and Xavier Rival.
\newblock The {ASTR\'EE} analyzer.
\newblock In {\em ESOP}, number 3444 in LNCS, pages 21--30, 2005.

\bibitem[Figueroa~del Cid(2000)]{FigueroaPhD}
Samuel~A. Figueroa~del Cid.
\newblock {\em A Rigorous Framework for Fully Supporting the {IEEE} Standard
  for Floating-Point Arithmetic in High-Level Programming Languages}.
\newblock PhD thesis, New York University, 2000.

\bibitem[Filli{\^a}tre and Boldo(2007)]{Filliatre_Boldo_ARITH18}
Jean-Christophe Filli{\^a}tre and Sylvie Boldo.
\newblock Formal verification of floating-point programs.
\newblock In {\em ARITH~18}. IEEE, 2007.

\bibitem[Fre(2005{\natexlab{a}})]{gcj}
{\em Documentation for \texttt{gcj} (\texttt{gcc} 4.1.1)}.
\newblock Free Software Foundation, 2005{\natexlab{a}}.

\bibitem[Fre(2005{\natexlab{b}})]{gcc}
{\em The {GNU} compiler collection}.
\newblock Free Software Foundation, 2005{\natexlab{b}}.

\bibitem[Fre(2001{\natexlab{a}})]{MPC750}
{\em {MPC750} RISC Microprocessor Family User's Manual}.
\newblock Freescale Semiconductor, Inc., December 2001{\natexlab{a}}.
\newblock MPC750UM/D.

\bibitem[Fre(2001{\natexlab{b}})]{PowerPC-32}
{\em Programming Environments Manual For 32-Bit Implementations of the
  {PowerPC} Architecture}.
\newblock Freescale Semiconductor, Inc., December 2001{\natexlab{b}}.
\newblock MPCFPE32B/AD.

\bibitem[Goldberg(1991)]{Goldberg91}
David Goldberg.
\newblock What every computer scientist should know about floating-point
  arithmetic.
\newblock {\em ACM Comput. Surv.}, 23\penalty0 (1):\penalty0 5--48, 1991.
\newblock ISSN 0360-0300.

\bibitem[Gosling et~al.(2005)Gosling, Joy, Steele, and
  Bracha]{Java_language_spec3}
James Gosling, Bill Joy, Guy Steele, and Gilad Bracha.
\newblock {\em The {Java} Language Specification}.
\newblock The Java Series. Addison-Wesley, third edition, 2005.

\bibitem[Gosling et~al.(1996)Gosling, Joy, and Steele]{JavaSpec1}
James Gosling, Bill Joy, and Guy~L. Steele.
\newblock {\em The {J}ava Language Specification}.
\newblock Addison Wesley, 1st edition, 1996.
\newblock ISBN 0201634511.

\bibitem[Gosling et~al.(2000)Gosling, Joy, Steele, and Bracha]{JavaSpec2}
James Gosling, Bill Joy, Guy~L. Steele, and Gilad Bracha.
\newblock {\em The {J}ava Language Specification}.
\newblock Addison Wesley, 2nd edition, 2000.
\newblock ISBN 0201310082.

\bibitem[Goubault(2001)]{Goubault_SAS01}
\'Eric Goubault.
\newblock Static analyses of floating-foint operations.
\newblock In {\em SAS}, volume 2126 of {\em LNCS}. Springer, 2001.

\bibitem[Hoare(1969)]{Hoare_1969}
C.~A.~R. Hoare.
\newblock An axiomatic basis for computer programming.
\newblock {\em Communications of the ACM}, 12\penalty0 (10):\penalty0 576--580,
  1969.
\newblock ISSN 0001-0782.

\bibitem[IEC(1989)]{IEC-60559}
{\em International standard -- binary floating-point arithmetic for
  microprocessor systems}.
\newblock IEC, 2nd edition, 1989.
\newblock IEC-60559.

\bibitem[IEE(1985)]{IEEE-754}
{\em IEEE standard for Binary floating-point arithmetic for microprocessor
  systems}.
\newblock IEEE, 1985.
\newblock ANSI/IEEE Std 754-1985.

\bibitem[Int(1997)]{Pentium-1}
{\em {Intel} Architecture Software Developer's Manual Volume 1: Basic
  Architecture}.
\newblock Intel Corp., 1997.
\newblock order number 243190.

\bibitem[Int(2005)]{IA32-1}
{\em IA-32 {Intel} Architecture Software Developer's Manual Volume 1: Basic
  Architecture}.
\newblock Intel Corp., September 2005.
\newblock order number 253665-017.

\bibitem[ISO(1999)]{C99}
{\em International standard -- Programming languages -- C}.
\newblock ISO/IEC, 1999.
\newblock standard 9899:1999.

\bibitem[{Java Grande Forum Panel}(1998)]{JavaGrandeSC98}
{Java Grande Forum Panel}.
\newblock Java grande forum report: Making java work for high-end computing,
  November 1998.

\bibitem[Kahan(1987)]{Kahan_branch_cuts}
William Kahan.
\newblock Branch cuts for complex elementary functions, or much ado about
  nothing's sign bit.
\newblock In A.Iserles and M.J.D. Powell, editors, {\em The State of the Art in
  Numerical Analysis}, pages 165--211, Oxford, 1987. Clarendon Press.

\bibitem[Kahan and Darcy(1998)]{KahanDarcy98}
William Kahan and Joseph~D. Darcy.
\newblock How {J}ava's floating-point hurts everyone everywhere.
\newblock Available online, March 1998.

\bibitem[Leroy et~al.(2005)]{OCaml}
Xavier Leroy et~al.
\newblock {\em The {O}bjective {C}aml system release 3.09: Documentation and
  user's manual}.
\newblock INRIA, 2005.

\bibitem[Lions et~al.(1996)]{Flight_501_report}
Jacques-Louis Lions et~al.
\newblock Ariane 5: flight 501 failure, report by the inquiry board.
\newblock Technical report, European Space Agency (ESA) and Centre national
  d'\'etudes spatiales (CNES), 1996.

\bibitem[Lynch et~al.(1995)Lynch, Ahmed, Schulte, Callaway, and
  Tisdale]{AMD_K5_transcendental}
T.~Lynch, A.~Ahmed, M.~Schulte, T.~Callaway, and R.~Tisdale.
\newblock The {K5} transcendental functions.
\newblock In {\em {ARITH}-12, 12th {IEEE} Symposium on Computer Arithmetic},
  page 163. IEEE, 1995.

\bibitem[Martel(2002{\natexlab{a}})]{Martel_ESOP02}
Matthieu Martel.
\newblock Propagation of roundoff errors in finite precision computations: a
  semantics approach.
\newblock In {\em ESOP}, number 2305 in LNCS. Springer, 2002{\natexlab{a}}.

\bibitem[Martel(2002{\natexlab{b}})]{Martel_SAS02}
Matthieu Martel.
\newblock Static analysis of the numerical stability of loops.
\newblock In {\em SAS}, number 2477 in LNCS. Springer, 2002{\natexlab{b}}.

\bibitem[Martel(2006)]{Martel_HOSC2006}
Matthieu Martel.
\newblock Semantics of roundoff error propagation in finite precision
  computations.
\newblock {\em Journal of Higher Order and Symbolic Computation}, 19\penalty0
  (1):\penalty0 7--30, 2006.

\bibitem[Min\'e(2001)]{Mine_AST01}
A{.} Min\'e.
\newblock The octagon abstract domain.
\newblock In {\em AST 2001 in WCRE 2001}, pages 310--319. IEEE CS Press,
  October 2001.

\bibitem[Min\'e(2004{\natexlab{a}})]{Mine_PhD}
Antoine Min\'e.
\newblock {\em Domaines numériques abstraits faiblement relationnels}.
\newblock PhD thesis, \'Ecole polytechnique, 2004{\natexlab{a}}.

\bibitem[Min\'e(2004{\natexlab{b}})]{Mine_ESOP04}
Antoine Min\'e.
\newblock Relational abstract domains for the detection of floating-point
  run-time errors.
\newblock In {\em ESOP}, volume 2986 of {\em LNCS}, pages 3--17. Springer,
  2004{\natexlab{b}}.

\bibitem[Muller(2005)]{MullerULP}
Jean-Michel Muller.
\newblock On the definition of ulp(x).
\newblock Technical Report 2005-09, \'Ecole normale sup\'erieure de Lyon -
  Laboratoire de l'Informatique du Parall\'elisme, 2005.

\bibitem[Rice(1953)]{Rice_1953}
H.~G. Rice.
\newblock Classes of recursively enumerable sets and their decision problems.
\newblock {\em Transactions of the American Mathematical Society}, 74\penalty0
  (2):\penalty0 358--366, March 1953.

\bibitem[Rival(2003)]{XR_VMCAI_2003}
X.~Rival.
\newblock Abstract interpretation-based certification of assembly code.
\newblock In L.~Zuck et~al., editors, {\em VMCAI}, volume 2575 of {\em LNCS},
  pages 41--55. Springer, January 2003.

\bibitem[Rogers(1987)]{Rogers}
Hartley Rogers.
\newblock {\em Theory of Recursive Functions and Effective Computability}.
\newblock MIT Press, 1987.

\bibitem[Steele and White(1990)]{SteeleWhite90}
Guy~L. Steele and Jon~L. White.
\newblock How to print floating-point numbers accurately.
\newblock In {\em PLDI}, pages 112--126. ACM, 1990.
\newblock ISBN 0-89791-364-7.

\bibitem[Sun(2001)]{Sun-numeric}
{\em Numerical Computation Guide}.
\newblock Sun Microsystems, 2001.

\bibitem[Weisstein(2005)]{ContinuedFraction}
Eric~W. Weisstein.
\newblock Continued fraction.
\newblock From MathWorld, 2005.
\newblock URL \url{http://mathworld.wolfram.com/ContinuedFraction.html}.

\bibitem[Winskel(1993)]{Winskel_1993}
Glynn Winskel.
\newblock {\em The Formal Semantics of Programming Languages: An Introduction}.
\newblock Foundations of Computing. MIT Press, 1993.
\newblock ISBN 0-262-23169-7.

\end{thebibliography}

\end{document}